\documentclass[preprint2]{aastex}
\usepackage{natbib}

\def\kms {\hbox{km\,s$^{-1}$}}
\def\ergs {\hbox{erg\,s$^{-1}$}}

\def\ccm {{\rm cm}^{-3}}
\def\gccm {{\rm g \ cm}^{-3}}
\newcommand{\e}[1]{\times 10^{#1}}
\newcommand{\iso}[1]{${}^{#1}$}

\newcommand{\wl}{$\lambda$ }
\newcommand{\msun}{$M_\odot$}
\newcommand{\wll}{$\lambda \lambda$ }

\newcommand{\myemail}{claes@astro.su.se}
\def\ergsm{\rm ~erg~s^{-1} cm^{-2}}
\def\Lya{{\rm Ly}\alpha}
\def\Ha{{\rm H}\alpha}
\def\Hb{{\rm H}\beta}

\begin{document}

\title{Late Spectral Evolution of the Ejecta and Reverse Shock in SN1987A}

\author{Claes Fransson\altaffilmark{1}, Josefin Larsson\altaffilmark{2}, Jason Spyromilio\altaffilmark{3}, Roger Chevalier\altaffilmark{4}, Per Gr\"oningsson\altaffilmark{1}, Anders Jerkstrand\altaffilmark{7,1}, Bruno Leibundgut\altaffilmark{3}, Richard McCray\altaffilmark{5}, Peter Challis\altaffilmark{6}, Robert P. Kirshner\altaffilmark{6}, Karina Kjaer\altaffilmark{3}, Peter Lundqvist\altaffilmark{1} and Jesper Sollerman\altaffilmark{1}}

 %\offprints{Claes Fransson, claes@astro.su.se}

\altaffiltext{1}{Department of Astronomy, The Oskar Klein Centre,
              Stockholm University, Alba Nova University Centre, SE-106 91 Stockholm, Sweden  \myemail}
   \altaffiltext{2} {KTH, Department of Physics, and the Oskar Klein
Centre, AlbaNova, SE-106 91 Stockholm, Sweden}
   \altaffiltext{3} { ESO, Karl-Schwarzschild-Strasse 2, 85748 Garching, Germany}
   \altaffiltext{4} {Department of Astronomy, University of Virginia, P.O. Box 400325, Charlottesville, VA 22904-4325, USA}
     \altaffiltext{5}  {JILA, University of
Colorado, Boulder, CO 80309Ð0440, USA.}
   \altaffiltext{6} {Harvard-Smithsonian Center for Astrophysics, 60 Garden Street, MS-19, Cambridge, MA 02138, USA. }
   \altaffiltext{7} {School of Mathematics and Physics, Queens University Belfast, Belfast BT7 1NN, UK}

    % \email{claes@astro.su.se}       	

  % \date{Draft 0}

\begin{abstract}
We present observations with VLT and HST of the broad
    emission lines from the inner ejecta and reverse shock of SN 1987A from
     1999 Feb. until 2012 Jan.  (days 4381 -- 9100 after explosion). We
    detect broad lines from $\Ha$, $\Hb$, Mg I], Na I, [O I], [Ca II]  and a feature at
  $\sim$ 9220 \AA. We identify the latter line with Mg II \wll 9218, 9244,
  which is most likely pumped by Ly$\alpha$
  fluorescence. $\Ha$, and $\Hb$ both have a centrally
  peaked component, extending to $\sim 4500$ \kms \
   and a very broad component extending to $\ga 11,000$ \kms,
  while the other lines have only the central component.  The low velocity component comes from unshocked
ejecta, heated mainly by X-rays from the circumstellar environment, whereas the broad component
comes from faster ejecta passing through the reverse shock, created by the collision with the circumstellar ring. 
The flux in $\Ha$
  from the reverse shock has increased by a factor of $4-6$ from
  2000 to 2007. After that there is a tendency of flattening of the
  light curve, similar to what may be seen in soft X-rays and in the  
  optical lines from the shocked ring. The core component seen in $\Ha$, [Ca II]
 and Mg II  has experienced a similar  increase, which is consistent with
  that found from HST photometry. 
The ring-like morphology of the ejecta is explained as a result of the X-ray 
  illumination, depositing energy outside of the core of the 
  ejecta. The energy deposition of the external 
  X-rays is calculated using explosion models for SN 1987A and we predict that the outer parts of the unshocked ejecta will 
  continue to brighten because of this.   We finally discuss evidence for 
  dust in the ejecta from line asymmetries.
  
 \end{abstract}


%\maketitle
%
%===============================================

\section{Introduction}
\label{sec_introd}
During the first decade after the explosion the spectrum of SN 1987A was dominated by
lines from newly synthesized metals in the inner ejecta, powered by the decay of \iso{56}Co, \iso{57}Co and
\iso{44}Ti. Now, more than twenty years after explosion the ejecta are
involved in an increasingly intense collision with the circumstellar
ring. This is seen in both optical/UV, radio and X-rays \citep[see
  e.g.,][for a review]{McCray2007}. In the optical the collision
manifests itself most clearly as a forest of increasingly bright
emission lines with velocities of $\sim 300$ \kms \
\ \citep{Pun2002,Groningsson2008a}. However, a number of
very broad lines, in particular Ly$\alpha$ and $\Ha$ with
velocities of $\ga 10^4$ \kms, are also evident in the spectrum.  Broad
H$\alpha$ and Ly$\alpha$ lines from the reverse shock were first seen
by \cite{Sonneborn1998}.  It is likely that most of that emission is
coming from collisional excitation of the neutral expanding H I by
shocked electrons, giving an extremely broad component
\citep{Michael1998b}. \cite{Michael1998a,Michael2003} found from
modeling of narrow slit observations with HST that this emission was
concentrated to a region of $\sim \pm 30\degr$ from the equatorial
ring, with much less emission from higher latitudes.

The evolution of the broad H$\alpha$ was discussed by \cite{Smith2005}
from a combination of observations with STIS and the Magellan 6.5 m
telescope. An interesting prediction was that the ionizing photons
from the reverse shock would pre-ionize the high velocity ejecta gas
and thereby quench the broad Ly$\alpha$ and $\Ha$ emission from the
reverse shock. Because of the limited signal to noise (S/N) and spectral resolution in the STIS spectra and
differences in the slit width, they could, however, not make any
firm statement on the evolution of the $\Ha$ flux.

Both the Ly$\alpha$ and H$\alpha$ evolution from 1999 to 2004 were discussed by
\cite{Heng2006} from STIS observations, who found an
increase in both the Ly$\alpha$ and H$\alpha$ fluxes by factors 
5.7 to 9.4 and 2 to 3, respectively. The larger increase of
Ly$\alpha$ was explained as a result of resonance back-scattering
of the Ly$\alpha$ photons. They also discussed what they
called the `interior emission', i.e., line emission. This could either originate in low
velocity ejecta or from charge transfer in the post-shock gas of slow
H II with high velocity H I. A careful analysis of this mechanism by
\cite{Heng2007}, however, found that this `interior emission',
corresponding to the `broad component' in Balmer dominated shocks,
was too strong to be explained by the charge transfer model. 

\cite{France2010} discussed the Ly$\alpha$ and H$\alpha$ emission based on STIS
observations from  2010. They find a further increase in the
Ly$\alpha$ and H$\alpha$ fluxes, for the latter by a factor of $\sim
1.7$ from 2004 to 2010. Most interesting, from the large
extent of the blue wing of Ly$\alpha$ they propose that Ly$\alpha$
photons from the hot-spots are boosted in energy by scattering against
the expanding ejecta. This works for Ly$\alpha$ since it is a
resonance line but not for H$\alpha$.

Recently, \citet[][in the following L11]{Larsson2010} found from 
photometry of the ejecta images from HST that while the flux from the 
ejecta decayed as expected from the \iso{44}Ti decay until 2001 (day $\sim$ 5000), 
it has increased since then. By 2010 the flux was $\sim 3$ times higher in the B
and R-bands than in 2001. While L11 mainly used the HST
photometry, the flux increase was supported by observations of the
flux of the [Ca II] \wll 7292, 7324 lines, that will be discussed in this
paper. In L11 it was proposed that the increase in the optical flux is caused 
by reprocessing of the X-ray and EUV emission from the ejecta - ring collision. 
This is further strengthened by the changing morphology of the ejecta as 
seen in the HST imaging, where Larsson et. al. (2012, in prep., in the following L13) find that 
most of the $\Ha$ emission from the inner ejecta has a ring-like morphology. 
This is in contrast to the emission in the 1.644 $\mu$m [Si I]/[Fe II] 
emission, which mainly comes from the core \citep[see also][]{Kjaer2010}. 

In this paper we discuss ground based high S/N observations with high
spectral resolution from the Very Large Telescope (VLT), complemented
by STIS observations from HST. A high S/N is especially
important in order to trace the faint line wings to high velocity and
also to detect additional, weaker lines in addition to the ones
discussed by \cite{Smith2005} and \cite{Heng2006}. In particular, we
discuss the evolution of the reverse shock in time and the evolution
of the lines from the inner regions of the ejecta. These are
particularly interesting to monitor since they may signal the
re-ionization of the inner ejecta by the hard radiation from the ring
collision. 

In Sect. \ref{sec_obs} we discuss the observations and reductions, and in 
Sect. \ref{sec_results} we detail the results. In Sect. \ref{sec_discuss} 
we discuss the implications for the reverse shock and for the emission from the ejecta. Summary and conclusions are given in Sect. \ref{sec_summary}.

\section{Observations}
\label{sec_obs}
\subsection{VLT observations}
\label{sec_vlt}

The ground based observations were performed with the VLT at
ESO, using the FORS2 and UVES instruments. SN 1987A has been monitored
regularly with these instruments since 1999. Table \ref{tab:obslog_uves} and Table
\ref{tab:obslog_fors} summarize the UVES and FORS2 observations. Primarily, the purpose has
been to follow the evolution of the narrow lines from the ring
collision. We have therefore mainly used the UVES high resolution
spectrograph. To maximize the spectral resolution we have chosen a
comparatively narrow slit, 0.8\arcsec \ wide.

\begin{figure}[!h]
\resizebox{\hsize}{!}{\includegraphics{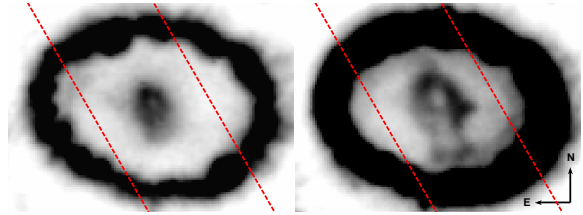}}
\caption{Slit position for the UVES observations overlaid on
  HST/WFPC2 F675W images from  2000 Nov. 13 (left) and   2009 Apr. 29
  (right). A slit width of 0.8\arcsec \ was used in all observations, except for the 1999 observation.}
\label{fig_slits}
\end{figure}

In most seasons, observations with seeing of order 0.6\arcsec \
were obtained, just enough to spatially resolve the ring into the
northern and southern parts \cite[see e.g.,][]{Groningsson2008a}. The position angle of the slit was
30\degr \ in all observations.  In Fig. \ref{fig_slits} we show the
slit orientations and widths for the UVES observations at two epochs. 

The FORS2 observations were obtained with a wide 1.6\arcsec \ slit and
are accompanied by a local spectrophotometric standard. To estimate slit
losses of the UVES spectra we have also obtained low resolution FORS2 observations with a narrow
slit (see Table \ref{tab:uves_fors})
Additionally, the slit acquisition images of the UVES spectrograph
have been inspected to ensure that the alignment of the instrument slit
to the supernova was consistent throughout the data set. This, as described in
\cite{Groningsson2008a}, gives us confidence that the error in the
absolute flux calibration is between not larger than 30\%. Furthermore, the UVES
observations employ a 12\arcsec \ long slit. This provides  a clean observation of H$\alpha$ emission from the LMC at the edges
of the slit, which
it is plausible to assume is invariant over the period of our observations
and can be expected to be extended with respect to the slit (therefore
insensitive to seeing).  The variation of the flux in the LMC lines is
well within 20\%. The uncertainties are
considered to be almost exclusively due to the position of the supernova
with respect to the slit and the seeing rather than throughput of the
instrument or the atmosphere.

The reductions of the FORS2 data followed classical long slit techniques using
the {\tt FIGARO} data reduction package. The UVES reductions used the
ESO pipeline. All wavelength settings were processed using both optimal
extraction and a manual extraction along the slit. For the H$\alpha$
and H$\beta$ lines the strong emission from the narrow circumstellar
lines poses a challenge for optimal extraction algorithms, and the
2-dimensional manual extraction and sky subtraction was preferred. The
difference between the two methods only affects the narrow components
which do not form part of this work. However, for the broad H$\alpha$
and H$\beta$ lines we have
used the robust 2-D extractions here. 

\begin{deluxetable}{l l c c c c c c c c}
\tabletypesize{\scriptsize}
\tablecaption{VLT/UVES observations of SN 1987A}
\tablewidth{0pt}
\tablehead{
\colhead{Epoch}& \colhead{Date}&\colhead{Days after }& \colhead{Setting} & \colhead{$\lambda$ range}&\colhead{Slit width} &\colhead{Resolution}&\colhead{Exposure}&\colhead{Seeing}&\colhead{Air mass}\\ 
&&\colhead{explosion$^{\mathrm{a}}$}&&\colhead{(nm)}&\colhead{(arcsec)}&\colhead{$(\lambda/\Delta\lambda$)}&\colhead{(s)}&\colhead{(arcsec)}
}
\startdata
1&1999 Oct 16&4618&346+580&303 -- 388&1.0&40,000&1,200&1.0&1.4\\
&&&&476 -- 684&&&&\\
&&&&&&&&\\
2&2000 Dec 10 -- 14&5040&346+580&303 -- 388&0.8&50,000&10,200&0.4--0.8&1.4\\
&&&&476 -- 684&&&&\\
&Dec 09 -- 10&5038&390+860&326 -- 445&&&9,360&0.4&1.4--1.6\\
&&&&660 -- 1060&&&&\\
&&&&&&&&\\
3&2002 Oct 06 -- Dec 14&5738&346+580&303 -- 388&0.8&50,000&10,200&0.7--1.0&1.5--1.6\\
&&&&476 -- 684&&&&\\
&Dec 13 - 14&5772&390+564&326 -- 445&&&9,360&0.4--1.1&1.4--1.5\\
&&&&458 -- 668&&&&\\
&Oct 04 - 05&5702&437+860&373 -- 499&&&9,360&0.4--1.1&1.4--1.5\\
&&&&660 -- 1060&&&&\\
&&&&&&&&\\
4&2005 Mar 21 - Apr 12&6611&346+580&303 -- 388&0.8&50,000&9,200&0.6--0.9&1.5--1.8\\
&&&&476 -- 684&&&&\\
&Apr 09 - 11&6621&437+860&373 -- 499&&&4,600&0.5&1.6--1.7\\
&&&&660 -- 1060&&&&\\
&&&&&&&&\\
5&2005 Nov 01&6826&346+580&303 -- 388&0.8&50,000&2,300&0.9&1.4\\
&&&&476 -- 684&&&\\
&Oct 20 - Nov 15&6825&437+860&373 -- 499&&&9,200&0.5--1.0&1.4--1.6\\
&&&&660 -- 1060&&&&\\
&&&&&&&&\\
6&2006 Oct 01 -- 21&7170&346+580&303 -- 388&0.8&50,000&9,000&0.5--0.9&1.4--1.5\\
&&&&476 -- 684&&&\\
&Oct 29 - Nov 15&7196&437+860&373 -- 499&&&9,000&0.5--1.0&1.4\\
&&&&660 -- 1060&&&&\\
&&&&&&&&\\
7&2007 Oct 23 - Nov 28 &7565&346+580&303 -- 388&0.8&50,000&11,250&0.8--1.4&1.4--1.6\\
&&&&476 -- 684&&&\\
&Oct 23&7547&437+860&373 -- 499&&&9,000&1.1--1.4&1.4--1.6\\
&&&&660 -- 1060&&&&\\
&&&&&&&&\\
8&2008 Nov 23 - 2009 Feb 08&7982&346+580&303 -- 388&0.8&50,000&9,000&0.8--1.0&1.4--1.5\\
&&&&476 -- 684&&&\\
&2009 Jan 09 -- 25&7998&437+860&373 -- 499&&&11,250&0.8--1.3&1.4--1.5\\
&&&&660 -- 1060&&&&\\
&&&&&&&&\\
9&2010  Nov 06 -- 15&8661&346+580&303 -- 388&0.8&50,000&9,000&0.7--1.1&1.4--1.7\\
&&&&476 -- 684&&&\\
&2010 Oct 19 -- Nov 15&8652&437+860&373 -- 499&&&11,250&0.6--1.1&1.4\\
&&&&660 -- 1060&&&&\\
&&&&&&&&\\
10&2011 Nov 06 -- Dec 03&9019&346+580&303 -- 388&0.8&50,000&9,000&0.8--1.1&1.4--1.5\\
&&&&476 -- 684&&&\\
&2011 Nov 05 -- Dec 04&9035&437+860&373 -- 499&&&9,000&0.6--1.4&1.4--1.7\\
&&&&660 -- 1060&&&&\\
\enddata
\label{tab:obslog_uves}
\begin{list}{}{}
 \item[$^{\mathrm{a}}$] Average epoch of spectrum since explosion, 1987 Feb. 23 .
 \end{list}
 \end{deluxetable}

\begin{deluxetable}{l l c c c c c c c c}
\tabletypesize{\scriptsize}
\tablecaption{VLT/FORS and HST/STIS observations of SN 1987A}
\tablewidth{0pt}
\tablehead{
\colhead{Epoch}& \colhead{Date}&\colhead{Days after }& \colhead{Setting} & \colhead{$\lambda$ range}&\colhead{Slit width} &\colhead{Resolution}&\colhead{Exposure}&\colhead{Seeing}&\colhead{Air mass}\\ 
&&\colhead{explosion$^{\mathrm{a}}$}&&\colhead{(nm)}&\colhead{(arcsec)}&\colhead{$(\lambda/\Delta\lambda$)}&\colhead{(s)}&\colhead{(arcsec)}
}
\startdata
STIS&1999 Feb 21 -- 27& 4384&750L&524 -- 636&$0.5$&666&$10,500$&&\\
STIS&1999 Aug 30 -- 31& 4571&750M&630 -- 687&$3 \times 0.1$&6,000&$3 \times 7,804$&&\\
&&&&&&&&\\
FORS1&2002 Dec 30&5788&600R&514 -- 730&0.70&1,660&5,400&0.7--0.9&1.4--1.6\\
FORS1&2002 Dec 30&5788&600B&336 -- 576&0.70&1,110&7,200&0.7--1.0&1.4\\
&&&&&&&&\\
STIS&2004 Jul 18 -- 23& 6358&750L&524 -- 636&$3 \times 0.2$&666&$3 \times 5,468$&&\\
&&&&&&&&\\
FORS2&2006 Nov 24&7213&600RI&512 -- 845&0.7&620&1,200&0.7&1.5\\
FORS2&2006 Nov 24&7213&600RI&512 -- 845&1.62&620&1,200&0.7&1.5\\
FORS2&2006 Dec 21&7240&1028Z&786 -- 962&1.62&1,580&1,320&0.8&1.4\\
FORS2&2006 Dec 21&7240&1200R&590 -- 740&1.62&1,320&1,320&0.9&1.4\\
FORS2&2006 Dec 21&7240&1200R&590 -- 740&0.70&3,060&1,380&1.1&1.4\\
FORS2&2006 Dec 21&7240&1400V&465 -- 596&1.62&1,300&1,380&0.7&1.4\\
&&&&&&&&\\
FORS2&2007 Nov 07&7561&600RI&537 -- 870&1.62&620&600&0.9--1.1&1.4\\
FORS2&2007 Nov 07&7561&1028Z&786 -- 962&1.62&1,580&1,740&1.0--1.2&1.4--1.5\\
FORS2&2007 Nov 07&7561&1200R&590 -- 740&1.62&1,320&600&1.3--1.4&1.4\\
FORS2&2007 Nov 07&7561&1200R&590 -- 740&0.70&3,060&600&0.9--1.0&1.4\\
FORS2&2007 Nov 07&7561&1400V&465 -- 596&1.62&1,300&900&1.0--1.2&1.4\\
&&&&&&&&\\
FORS2&2008 Nov 04 -- Dec 26&7950&600RI&537 -- 870&1.62&620&1,200&0.9--1.0&1.4\\
FORS2&2008 Nov 28 -- Dec 03&7951&1400V&465 -- 596&1.62&1,300&1,800&0.6--1.0&1.4--1.5\\
FORS2&2008 Dec 26&7976&1200R&590 -- 740&0.70&3,060&600&0.9&1.4\\
FORS2&2008 Dec 26&7976&1200R&590 -- 740&1.62&1,320&600&0.8--1.0&1.4\\
FORS2&2008 Dec 26&7976&1028Z&786 -- 962&1.62&1,580&1,740&0.8--0.9&1.4\\
&&&&&&&&\\
STIS&2010 Jan 01& 8378&750L&524 -- 636&$0.2$&666&$14,200$&&\\
&&&&&&&&\\
FORS2&2011 Nov 25&9040&600RI&537 -- 870&1.62&620&1,200&0.9--1.0&1.41\\
FORS2&2011 Dec 03&9049&1400V&465 -- 596&1.62&1,300&1,800&0.6--1.0&1.55\\
FORS2&2012 Jan 14&9090&1200R&590 -- 740&1.62&1,320&1,400&0.8--1.0&1.41\\
FORS2&2012 Jan 14&9090&1028Z&786 -- 962&1.62&1,580&3,480&0.6--0.9&1.41\\
FORS2&2012 Jan 23&9099&1200R&590 -- 740&0.70&3,060&1,200&0.9&1.55\\
\enddata
\label{tab:obslog_fors}
\begin{list}{}{}
 \item[$^{\mathrm{a}}$] Average epoch of spectrum since explosion, 1987 Feb. 23.
 \end{list}
\end{deluxetable}

\begin{figure}[!h]
\resizebox{\hsize}{!}{\includegraphics{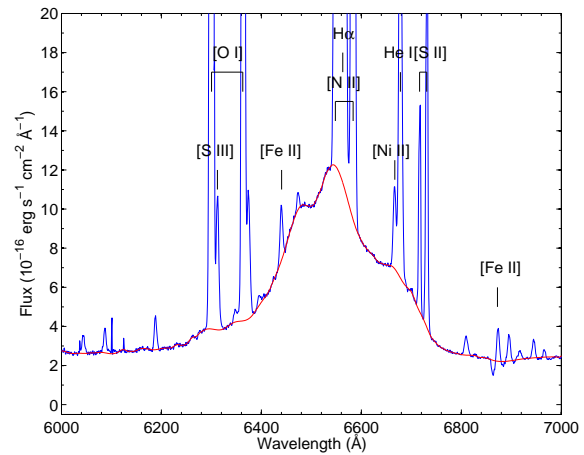}}
\caption{H$\alpha$ profile before (blue) and after (red line) subtraction of the narrow
  and intermediate velocity lines for the FORS2 1200R spectrum (slit
  width 1.62\arcsec) at 7976 days. }
\label{fig2}
\end{figure}

In this paper we concentrate on the evolution of the broad lines from
the ejecta and reverse shock. To show this more clearly we 
subtract the narrow and intermediate velocity lines by a cubic spline
interpolation between the extremes of the narrow
lines. Figure \ref{fig2} shows the result before and after subtraction
for $\Ha$ for the day 7976 FORS spectrum. Even if the resulting line
has a smooth shape, one should note that especially the peak of this
line is uncertain because of the narrow line contamination. In some cases
comparison with HST spectra helps in this respect, even if their
spectral resolution is much lower (Sect. \ref{sec_stis}).

The UVES slit only covers part of the ejecta and this affects the flux and the
line profiles we observe. The major axis of the ring
is 1.6\arcsec \ and the minor axis 1.1\arcsec, so only part of the
outer ejecta, which now fills most of the ring, is within the UVES
slit (Fig. \ref{fig_slits}).  Because most of the flux in the high 
velocity wings come from the
central parts of the projected image along the line of sight (LOS), these
are likely to be less affected than the low velocity parts of the
line. These come both from the supernova core and the outer parts of the
projected image. Part of the latter emission may be missed by using
a narrow slit.

\begin{figure}[!h]
\resizebox{\hsize}{!}{\includegraphics{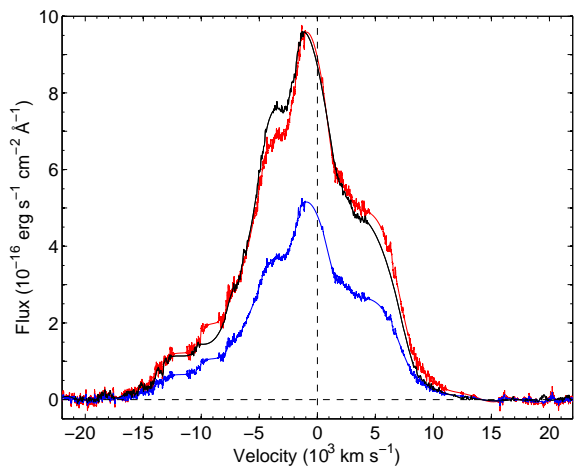}}
\caption{Comparison of $\Ha$ taken with FORS2 with a 1.6\arcsec \ slit
   and UVES with a 0.8\arcsec \ slit at 7976 days and 7982 days,
   respectively. The lower, blue spectrum is the original UVES
   spectrum, while the black spectrum is from FORS2. The red spectrum
   is the UVES spectrum multiplied by a factor 1.86. 
   Most of the remaining
   difference between the UVES and FORS2 spectra come from
   velocities $\sim  5000$ \kms, characteristic of the
  reverse shock close to the ring plane and outside the UVES slit.
  }
\label{fig1}
\end{figure}

Fig. \ref{fig1} shows a comparison between one of the slit integrated
spectra obtained with UVES with the 0.8\arcsec \ slit and the FORS2 spectrum with
the 1.6\arcsec \ slit for $\Ha$. The FORS2
slit covers most of the ejecta, while the UVES slit only
covers the central fraction of it.
The exact fraction of the total flux depends on the seeing and the
spatial origin of the emission. In Fig.  \ref{fig1} we have subtracted
the continuum, which mainly comes from the ring collision (Sect. \ref{sec_results}). Most of the
difference in the FORS2 and UVES line profiles comes from the parts of
the ejecta outside the slit, i.e., in the NW and SE directions. In the
ring plane, where most of the emission from the reverse shock
originates, this corresponds to velocities in the range  -5000 -- +5000
\kms, while higher velocities, coming mainly from material expanding
in our direction, should fall within the narrow UVES slit. 

In Table \ref{tab:uves_fors} we
give the ratio of the H$\alpha$ fluxes with the different
slit widths at the epochs when we have observations with both instruments. This indicates that the fraction lost
stays nearly constant with time and is close to proportional to the width
of the slit. 
We note that inner ejecta, moving across the line of sight at $\la 4000$ \kms \ 
for 20 years will still be within 0.33 \arcsec \ of the remnant's center.
 Although slightly spread out by the seeing, most of this emission should then fall within the 0.8\arcsec \
slit of UVES. Most observations were made with a seeing $\la 0.8\arcsec$ and the comparison with the FORS observations indicate that the slit losses are small for the core component. (See Fig. \ref{fig_core_rev}, below.)

\begin{deluxetable}{l c  c}
\tablecaption{Comparison of $\Ha$ fluxes measured with FORS2 and UVES with different slits.}
\tablewidth{0pt}
\tablehead{
\colhead{Days after explosion}&\colhead{FORS2(1.6\arcsec)/UVES(0.8\arcsec)}&\colhead{UVES(0.8\arcsec)/FORS(0.7\arcsec)}\\ 
\colhead{FORS / UVES}&&}
\startdata
7240  / 7170&1.9&0.96\\
7561 /  7565&1.8&1.18\\
7976 / 7982&1.9&1.24\\
9090  / 9019&2.0&\\
\enddata
\label{tab:uves_fors}
\end{deluxetable}

\subsection{HST STIS Observations}
\label{sec_stis_obs}
The HST STIS observations used in this paper were carried out in 1999 (day 
4381, G750L grating, and days 4571-4572, G750M grating), 2004 (days 6355-6360, 
G750L grating) and 2010 (day 8378, G750L grating). The G750L grating has a 
spectral resoltion of $\sim 450\ $\kms \ and covers the
wavelength interval between  5240 -- 10270 \AA, which includes H$\alpha$ 
and [Ca II]. The spectral resolution of the G750M grating is significantly 
better ($\sim 50\ $\kms), but in this observation the wavelength coverage was 
reduced to 6295 -- 6867 \AA, which means that only the H$\alpha$ line is included. 

For the  1999 observation G750L a wide slit of $0.5$\arcsec \  was used 
\citep{Michael2003} at a position angle 25.6\degr.  In addition,  
spectra with the G750M grism with three parallel slits of $0.1$\arcsec  \ width at a position 
angle 27\degr \ were taken. For the 2004 observation \citep{Heng2006} 
a $0.2$\arcsec \ slit with a position angle of 0\degr \ was
placed in three different locations, thus covering all of the inner
ejecta as shown in Fig.~\ref{fig3f}. Finally, for the 2010 observation 
\citep{France2010} one 0.2 \arcsec \ slit with position angle 0\degr \ was 
used, covering only the central part of the ejecta perpendicular to the slit. 
 Details of all observations are summarized in Table \ref{tab:obslog_fors} 
and in more detail in Table 2 in L13 and we 
refer the reader to that paper for details regarding the data reduction.

\section{Results}
\label{sec_results}
\begin{figure*}
\centering
\includegraphics[width=16cm]{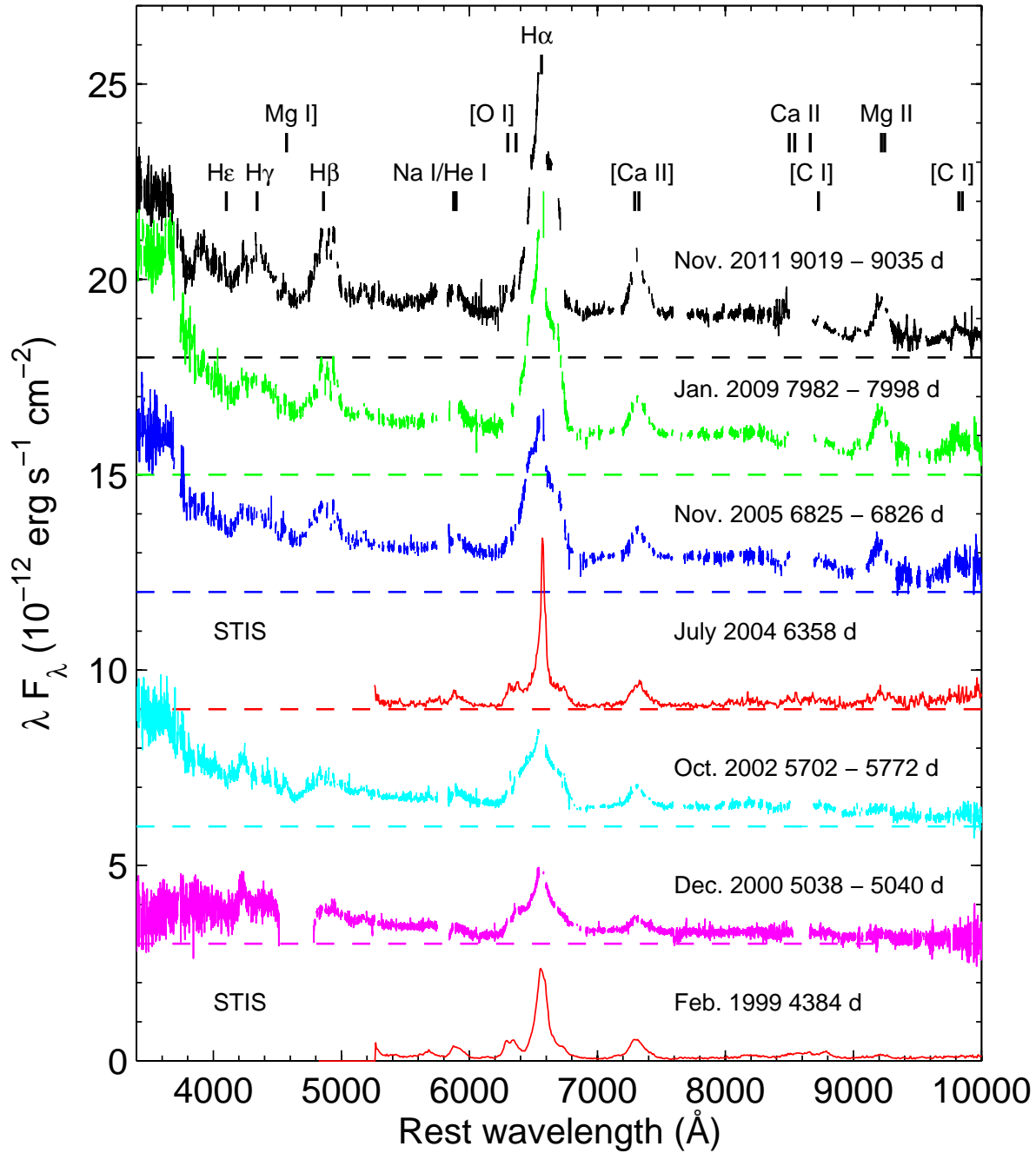}
\caption{Compilation of UVES and STIS spectra from the ejecta and reverse shock, with
  the narrow lines from the ring subtracted. Each spectrum has
  been been shifted upwards by $3 \e{-12} \ergsm$ \ relative to the
  previous. The dashed lines give the zero
flux level for each date. Both
  H$\beta$ and H$\gamma$ are blended with Fe I-II lines. The gaps in the
  UVES spectra are either due to gaps between the grisms or regions
  severely contaminated by lines from the ring. 
  All spectra have been de-reddened with E$_{B-V} = 0.19$ mag. 
  This figure, unlike the others, is displayed in $\lambda F_\lambda$ for
clarity. }
\label{fig_ejecta_spectra}
\end{figure*}

In Fig. \ref{fig_ejecta_spectra} we show a subset of the full UVES spectra from Dec. 2002 
until  2011 Nov., together with the STIS spectra from
 1999 Feb. and  2004 Jul. (Sects. \ref{sec_stis_obs} and \ref{sec_stis}). 
Also here we have subtracted the many narrow lines originating from the unshocked
and the shocked ring from the UVES spectra. These spectra have all been de-reddened by 
E$_{B-V} = 0.19$ mag using the \cite{Cardelli1989} reddening law. To show the relative energy 
distribution we plot in this figure $\lambda F_\lambda$ .

Most of the rising continuum towards the UV
is Paschen and two-photon emission from the ring collision, which is
apparent when we compare the continuum with that in the STIS
observations, which isolate the ejecta better
(Sect. \ref{sec_stis}). The Balmer continuum below 3646 \AA \ from the same source is also
prominent.

Superimposed on this are several broad lines, in particular $\Ha,
\Hb$, Mg I] \wl 4571 and [Ca II] \wll 7292, 7324. We identify an
  interesting feature at $\sim$ 9220 \AA \ as emission by Mg II \wll 9218, 9244. In the
  blue wing of $\Ha$ there is most likely also a contribution from [O
    I] \wll 6300, 6364.  We discuss these and some weaker features
  further in Sect. \ref{sec_FeII}.

We have fit the total continuum spectrum with
the sum of free-bound and two-photon continua from H I and He I
\citep[e.g.,][]{Ercolano2006}.  The two-photon contribution will be suppressed at
higher densities, and its relative contribution will therefore depend on
the density. To include this effect we have employed a model atom for H I similar
to that used in \cite{Kozma1998I}. 

Our best fits correspond to a
temperature of $\sim 3 \e4$ K and a density of $\sim 10^6 \ \ccm$,
although the density could be higher without affecting the results
appreciably. These numbers are reasonable for the conditions in the
shocked gas of the ring. 

The first thing to note from Fig. \ref{fig_ejecta_spectra} is the
increase in the flux of all broad lines during this period. A good
example is the [Ca II] \wll 7292, 7324 doublet, which shows a steady
increase in the flux. We also note that the strength of the [Ca II]
lines in the STIS spectrum are in between the UVES 2002 Oct.  (day 5702) and  2005  Nov. (day 6825)
fluxes, which shows that we include most of the flux from the central
ejecta also in the UVES observations, despite the narrow slit. There
is, however, a large difference in the line profile of the H I lines between the UVES and STIS observations,
with much stronger 'shoulders' in the ground based observations. This
is caused by the larger extraction region along the slit for the UVES spectra compared to the STIS spectra (see Sect. \ref{sec_stis}). This results in a larger contribution from the reverse shock in the UVES spectra,

Because the hydrogen lines and the other lines differ substantially in
their properties and origin we discuss the results of these
separately.

\subsection{Hydrogen lines}
\label{sec_hydrogen}
\begin{figure}[!h]
 \resizebox{\hsize}{!}{\includegraphics{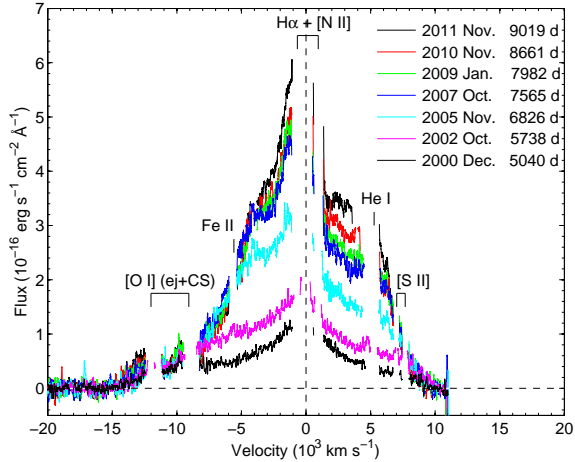}}
\caption{Evolution of the H$\alpha$ profile from 2000 to 2011 from the
  0.8\arcsec \  slit UVES observations. The positions of the most
  important narrow lines from the ring collision, which have been
  blocked out, are marked. The continuum, defined by the red and blue sides outside of H$\alpha$, has been subtracted for each spectrum.
  The blue wing above $\sim 8000$ \kms \ is dominated by a blend of [O I] \wll 6300, 6364 from the inner ejecta and the shocked ring. }
\label{fig4}
\end{figure}

Figure \ref{fig4} shows the evolution of the slit integrated $\Ha$
line profile with time from  2000 Dec. to   2011 Nov. from our UVES
observations. In addition to removing the narrow lines, we have here
subtracted the continuum level  by simply subtracting a constant flux for each date. Because of the slow variation of the continuum level this is a reasonable approximation, without introducing additional parameters.

\begin{figure}[!h]
\resizebox{\hsize}{!}{\includegraphics{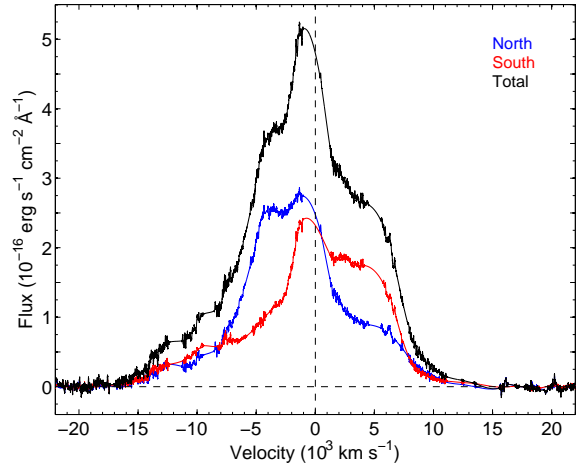}}
\caption{Comparison of the $\Ha$ line profiles for the northern and southern parts of H$\alpha$ at 7982 days post explosion.}
\label{fig3}
\end{figure}

Extracting different regions along the slit we can separate the northern and southern parts of the ejecta
as shown in Fig. \ref{fig3}. The northern component is clearly blueshifted, while the southern is redshifted. This was found already by  \cite{Smith2005}, and is discussed in detail in L13. 
We have also measured the evolution of the flux in the two regions, and find this to be very similar, with roughly equal fluxes in the two.

\begin{figure}[!h]
\resizebox{\hsize}{!}{\includegraphics{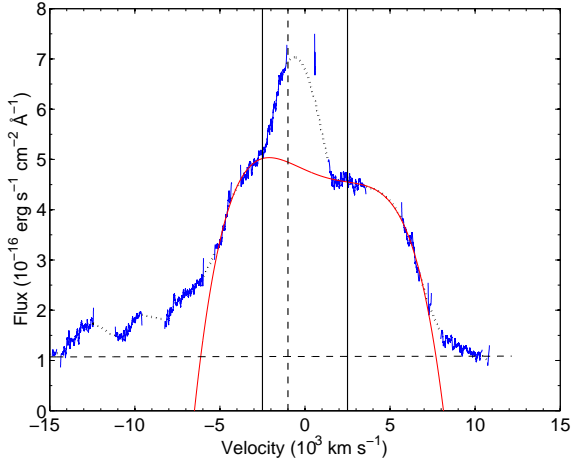}}
\caption{Illustration of the flux determination of the core component for the Dec. 2011 UVES spectrum. The solid, blue line shows the (smoothed) spectrum with the narrow lines removed. The dotted line in the gaps is the spline interpolation to the spectrum in the gaps. The solid, red line is a polynomial fit to the reverse shock between $-5000$ \kms \ and $+7000$ \kms, used for the calculation of the core flux. The vertical solid lines mark the $\pm 2500$ \kms \ velocity and the dashed, vertical line the $- 1000$ \kms \ velocity. The horizontal, dashed line shows the continuum level. The flux of the reverse shock is calculated between $- 8000$ \kms \  and $+ 10,000$ \kms, with the core component subtracted.}
\label{fig_core_split}
\end{figure}

From Figs. \ref{fig1} - \ref{fig4} one can distinguish two components to the line
profile, as was done  in  \cite{Smith2005}. One velocity
component reaching $\sim 4500$ \kms \ from the core, and one high velocity
component reaching velocities $\ga 10,000$ \kms. These evolve fairly
independently of each other, and to determine the flux in each of these
components we have assumed the broad component to include the part of
the line profile with velocity higher than $\sim 2500$ \kms \ on the blue
and red sides. 
Between these velocities we make
a polynomial interpolation between the red and blue parts of the profile
of the broad component, and assume that the flux above this represents
the contribution from the core and that below from the reverse
shock (see Fig. \ref{fig_core_split}  for the Dec. 2011 (day 9019) spectrum). This is a reasonable procedure since a radially thin shell, 
a fair approximation to the geometry of the reverse shock,
should have a flat line profile \citep[see e.g.,][for simulations with different geometries]{Michael2003}.

However, given the limitations of ground based observations, there may be substantial systematic errors in the fluxes determined. This applies especially to the core component, where the interpolation of the line profile in the red part of the line is uncertain. Also the division between the reverse shock and the core component introduces an additional uncertainty in the core flux. The much higher contribution to the flux by the reverse shock implies that it is less affected by this procedure. 

\begin{figure}[!h]
\resizebox{\hsize}{!}{\includegraphics{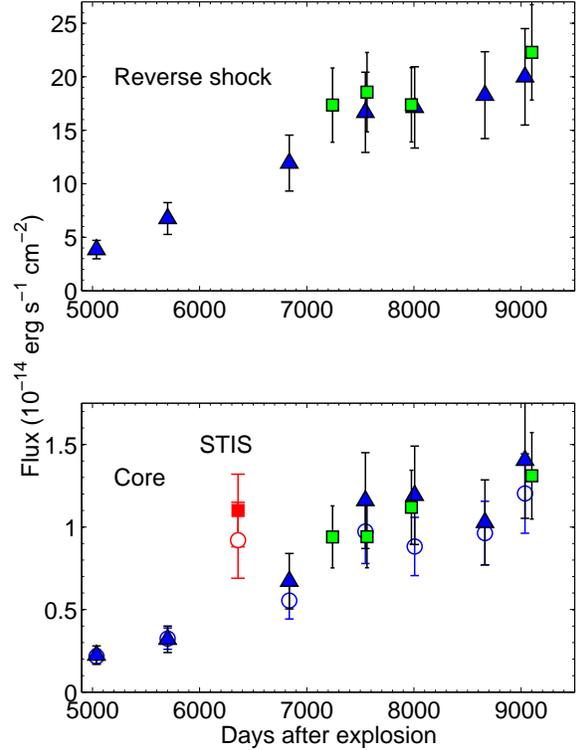}}
\caption{Upper panel: Evolution of the H$\alpha$ flux from  the reverse shock. 
Filled triangles show the flux from UVES, while filled squares are from FORS.  
 The reverse shock fluxes from UVES have been been multiplied by a factor 2.0 to compensate for the narrow slit. Lower panel: Same for the flux from the core. 
Filled triangles  show the UVES flux between $\pm 2500$ \kms, including the interpolated part of the line (see Fig. \ref{fig_core_split}), while the open circles are the fluxes from the part of the line unaffected by the narrow lines between $- 2500$ \kms \ and $- 1000$ \kms . These have been scaled by a factor of 4.0 for a comparison with the interpolated total core flux. 
The red points show the corresponding values measured with STIS. 
}
\label{fig_core_rev}
\end{figure}

In Fig.  \ref{fig_core_rev} we show the resulting light curves for the
core component and the
reverse shock component. Note that the UVES slit only covers 0.8\arcsec \
of the ejecta and to get the total flux from the reverse shock we therefore 
multiply these by the correction factor shown in Table
\ref{tab:uves_fors}. The flux from the core component should fall
mainly within the slit, although part of this may also fall outside due to the broadening of the PSF because of the seeing. We have in the same way determined the fluxes from the dates where FORS observations are available. This is an important check because the 1.6\arcsec \ slit should cover the full ejecta and reverse shock. This is shown as the filled squares in Fig. \ref{fig_core_rev}. 

As an additional check of the uncertainty introduced by the interpolation we also determine the core flux in the part of the line between $-2500$ \kms \ and $- 1000$ \kms \ not affected by the narrow lines, shown as open circles in Fig.  \ref{fig_core_rev} .

\begin{figure}[!h]
%\resizebox{\hsize}{!}{\includegraphics{./flux_vs_time_vs_vel2_cf_v2.eps}}
\resizebox{\hsize}{!}{\includegraphics{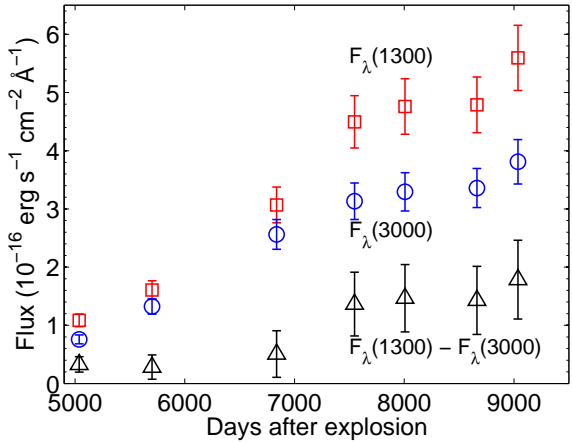}}
\caption{Evolution of the  monochromatic, continuum subtracted H$\alpha$ flux at -1300 \kms   \  (squares) and at -3000 \kms \ (circles)  and their difference (triangles). }
\label{fig5b}
\end{figure}

A third way of determining the evolution is to plot the monochromatic flux at
velocities characteristic of the central ejecta and reverse shock,
respectively.  In Fig. \ref{fig5b} we show the flux at -1300 \kms,
which is as close to the peak as one can safely trace H$\alpha$
without interference with the narrow lines, and at -3000 \kms, which
is at the plateau, as well as their difference. We believe that, although this only gives the monochromatic flux at these velocities, it gives a more accurate representation of the flux evolution than the total flux. 

For the reverse shock we see that, including the correction for the slit width, there is good agreement between the UVES and FORS fluxes, which shows that the more complete UVES set gives a good representation of the time evolution. 

The core fluxes are considerably more uncertain, both in terms of the absolute flux and time evolution. 
In Sect. \ref{sec_stis} we estimate the flux from the core from high spatial resolution STIS spectra. We there find a flux nearly a factor of two higher, which 
gives an estimate of the systematic errors in the ground based estimates, mainly caused by the contamination by the narrow lines. The fact that the total interpolated flux and the directly observable flux between $-2500$ to  $-1000 $ \kms show a very similar increase, however, indicates that the {\it relative} flux evolution is reasonably accurate. 
We discuss the implications of the light curves further in Sect. \ref{sec_rev_shock}.

We can compare the H$\alpha$ flux with that determined by \cite{Smith2005} on
2005 Feb. 25 (day 6577). On this day they find a total flux from
the reverse shock component of $1.37(\pm 0.15) \e{-13}  \ \ergs \rm
cm^{-2}$. This is approximately a factor of two higher than what we
determine from UVES for  2005 Apr. 
The slit they use is similar to ours, 0.8\arcsec, and the seeing is
also similar. The position angle is different from ours,
P.A. $-10$\degr, compared to 30\degr, and may affect the flux somewhat. However, they calibrate their flux
with the part of the HST F658N narrowband image of the ring that falls
within the slit. This should give a reasonable calibration, which also
compensates for the fraction of the light from this region that  is scattered outside the slit because of the
seeing in
the ground based image. In addition, they add a correction from the HST image for the
fraction of the flux which falls outside of the slit. 
After the correction applied to our data for the UVES slit from Table \ref{tab:uves_fors}, we find 
a fair agreement between our flux and that of
\cite{Smith2005}. To compare the FORS2 \& UVES data  
we have applied this correction
to our light curve. Although this does not include all flux, it does
result in a consistent light curve. 
The total flux in the reverse shock at each time can be found by
multiplying the flux in Figure \ref{fig_core_rev}  by the factor in Table  \ref{tab:uves_fors}.

The maximum velocity of $\Ha$ is not well defined on either the red or the blue side. On the blue side it is difficult to establish the
true extent of the line due to blending with the
broad [O I] \wll 6300, 6364 lines from the ejecta around 9096 \kms \ and
12,022 \kms, respectively. On the red side narrow and intermediate velocity lines from He I \wl 6855 and Fe II \wl 6872, interfere. We can therefore only establish lower limits to the maximum expansion velocities on each side, $\sim 9000$ \kms \ on the blue side while on the red side the line extends at least to 11,000 \kms, where the line is still substantially above the 'continuum' level, and could extend to considerably higher velocities, up to $\sim 13,000$ \kms.

Because of the difficulty in determining the exact continuum level the evolution of the flux at velocities $\ga 10,000$ \kms \ is uncertain. At lower velocities the flux is increasing monotonously up to $\sim 7500$ days, after which it flattens. 
The gaps in the line profile at high velocities make it difficult to establish whether the maximum velocity is 
affected by the interaction with the external medium.

The maximum velocity can be compared to the width
of the red wing of $\Ha$ at 7.87 years, which \cite{Chugai1997} found to
extend to $\sim 10,800$ \kms. Although the S/N in their observations was 
considerably lower, our velocity measurements are consistent with
theirs. \cite{Michael2003} report diffuse $\Ha$ emission out to $\sim
\pm 12,000$ \kms \ from the central direction of the ejecta in their
Oct. 1999 observation, presumably the same emission as is giving rise
to the high velocity wing in our observations. This is also supported
by the fact that this part of the line is similar in the northern
and southern part of the spectrum (Fig. \ref{fig3}), as expected from
a direction in the projected center. 
As we show in Sect. \ref{sec_stis}, also the STIS 2004 and 2010 spectra give a velocity close to this.

Most of the change in flux of H$\alpha$ takes place at low
and intermediate velocities.  A clear break in the profile can be seen  at $\pm 4000$ \kms (Fig. \ref{fig4}). 
Above this velocity the line profile shows a gradual decrease to the maximum velocity. 
The main change in flux of the UVES profiles occurs below $\sim 6000 \  \kms$ for both the blue and red wings, as can be seen in Figs. \ref{fig4} and \ref{fig5b}. At higher velocities the profile is nearly unchanged. %The same is true for the red side. 

\begin{figure}[!h]
\resizebox{\hsize}{!}{\includegraphics{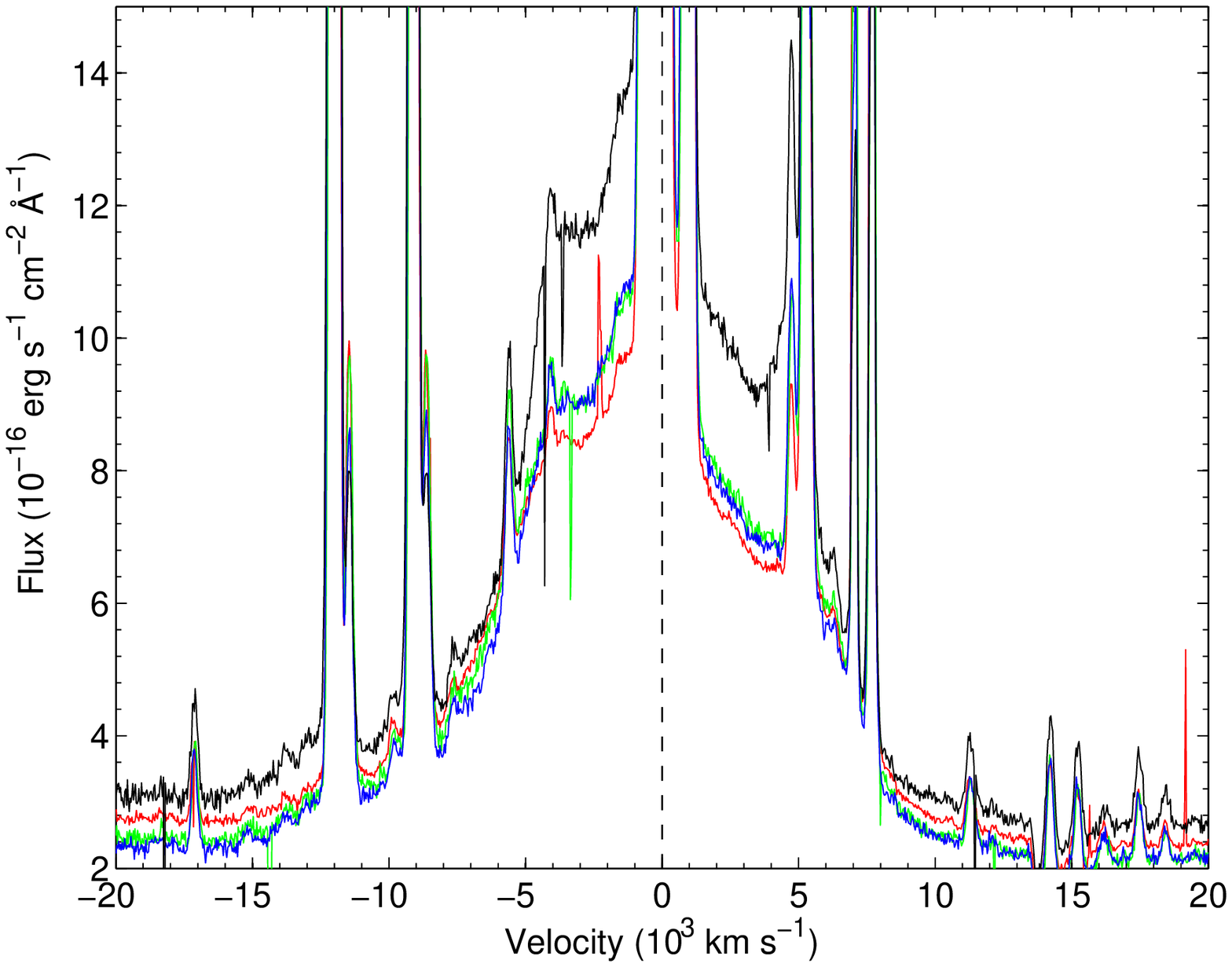}}
\resizebox{\hsize}{!}{\includegraphics{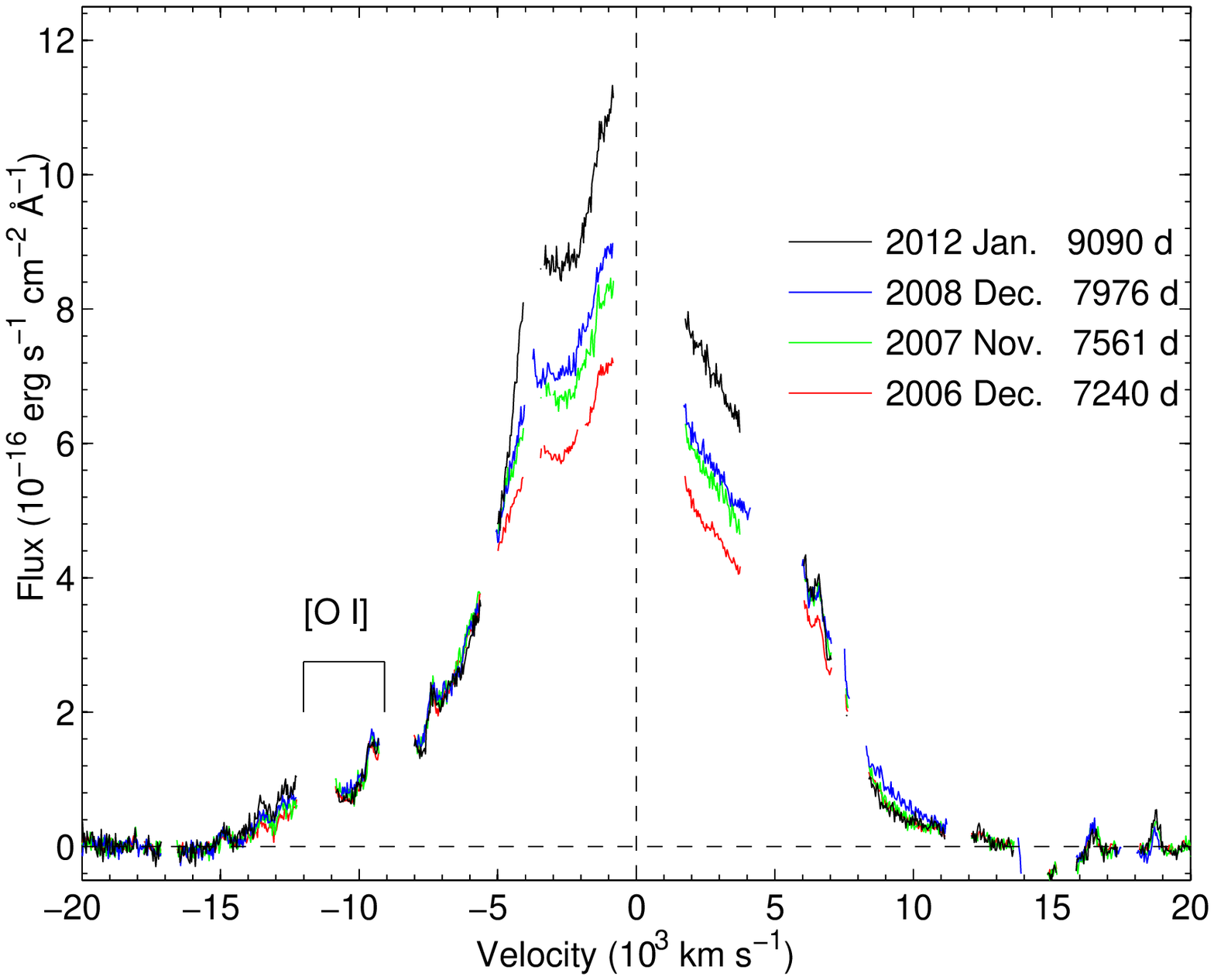}}
\caption{Comparison of H$\alpha$ profiles from FORS2 with a 1.6
 \arcsec \ slit from 2006 Dec. (day 7240) to  2012  Jan. (day
  9090) (bottom to top). The upper panel shows the original spectra, while the lower plot shows spectra with the narrow lines removed and the continuum subtracted. Note that it is mainly the part of the line core below $\sim 5000$
  \kms \ which has increased in flux during this period. The position of the [O I]  \wll 6300, 6364 doublet from the core, which influences the blue wing of the H$\alpha$ line, has been marked.}
\label{fig4a}
\end{figure}

This can be investigated in somewhat more detail from the FORS spectra in Fig.  \ref{fig4a}. When comparing the spectra from days 7240 and 7976 we see that the increase in flux only occurs for velocities below $\sim 5000$ \kms. If we now compare the spectra from days 7976 and 9090 the change only occurs for velocities below $\sim 4600$ \kms. Although these velocities are uncertain, and depend on e.g., the exact continuum level, we therefore note that the  velocity below which the flux increases is getting smaller with time. 

Below $\sim 2500 \  \kms$ there is a linear rise of the profile with time, which is likely to be caused by a superposition of the core component and the flat reverse shock profile. This can clearly be seen in the STIS observations discussed in Sect. \ref{sec_stis}. On the red side no flat section is seen, which is  explained by the fact that the region between $3300 - 6000 \ \kms$ is blocked by lines from the ring. We also note that there is an asymmetry between the red and blue sides, with a considerably lower flux on the red (southern) side.

When comparing the line profiles from the southern and northern parts
(Fig. \ref{fig3}) it is clear that the blue wing of the 'flat' part of
the line originates from the northern half of the ejecta/ring, while
the opposite is the case for the red part, as is expected from the
geometry. The higher
flux from the NE side of the debris, which produces the blue-shifted side
of the line profile, is consistent with the stronger ejecta-ring
interaction seen on that side  \citep[e.g.,][]{Racusin2009}.

\begin{figure}[!h]
\resizebox{\hsize}{!}{\includegraphics{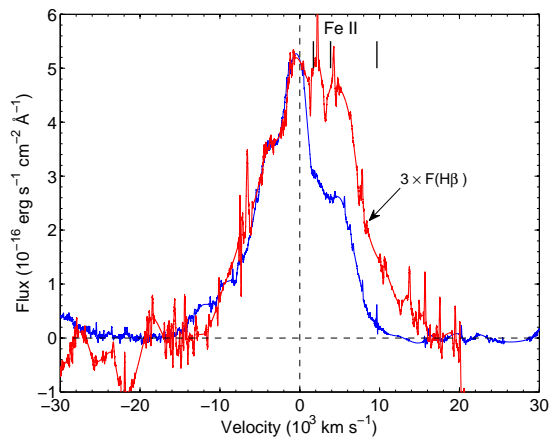}}
\caption{Comparison of the $\Ha$ (blue) and $\Hb$ (red) UVES profiles for day 7982 -- 7998. The
  $\Hb$ flux has been scaled by a factor 3.0 compared to $\Ha$. The
  red wing is probably blended with three Fe II lines, marked with
  their positions in velocity.}
\label{fig7b}
\end{figure}

Figure \ref{fig7b} shows $\Hb$ together with $\Ha$ for days 7982 -- 7998. While
the continuum level of $\Ha$ is well determined, $\Hb$ is in a crowded
region and the continuum is more difficult to define, which introduces some
uncertainty. By scaling the $\Hb$ flux by a factor 3.0, as in
Fig. \ref{fig7b}, we find good agreement between the two lines for the
blue wing. The red wing of $\Hb$ is, however, considerably stronger
compared to $\Ha$. This is most likely a result of blending with other
lines. In their 1995 observations \cite{Chugai1997} identify two Fe II
lines at 4889 \AA \ and 5018 \AA. In addition, the `bump' at $\sim
4000$ \kms \ can be identified with the strong Fe II a 6S -- z
6P${}^o$ \wl 4924 transition. The velocities of these lines are
marked in Fig. \ref{fig7b}. We also note the lower velocity of the
blue wing of $\Hb$ compared to $\Ha$, which confirms that this part of `$\Ha$' is caused by blending with
the [O I] \wll 6300, 6364 lines. The lines therefore
complement each other in filling several of the `gaps' caused by the
lines from the ring collision, as well as blending with other ejecta
lines.

Including a correction for reddening, the observed $\Ha/\Hb$ ratio of
3.0 corresponds to $\Ha/\Hb = 2.5$. There is no indication that this
ratio varies over the line profile. The measured ratio is, however, sensitive
to the chosen level of the continuum, which is especially problematic for
$\Hb$, where there are few regions free from other lines (see
Fig. \ref{fig7b}). Given this uncertainty, our line ratio is 
similar to that found by \cite{Chugai1997} at 7.87 years,
$\Ha/\Hb = 3.8$.  At that epoch the flux was, however, dominated by
the inner ejecta, with little contribution from the reverse shock. Because the emission from the reverse shock and from
the ejecta are produced by different mechanisms, there is no reason to
expect these ratios to remain the same (Sect. \ref{sec_rev_shock}). 
This ratio can be compared to modeling of the Balmer emission in supernova remnants, where  \cite{Chevalier1980} find an H$\alpha$/H$\beta $ ratio of 3 -- 5 depending on
   the optical depth in the Lyman lines.  The uncertainty in the value determined here is large enough that it is unclear if there is any diagnostic value in this ratio.

In addition to these lines one can in Fig. \ref{fig_ejecta_spectra} identify
a strong line coinciding with H$\gamma$, which is clearly blended with other lines.  Based on
the observations in \cite{Chugai1997} and modeling in \cite{Jerkstrand2011} these are mainly  Fe I, and in some cases also Fe II,
lines. The Ca I \wl 4226 line, found by \cite{Jerkstrand2011} to be strong in the 8 year spectrum, may also be present. Higher members of the Balmer series may  contribute below $\sim 4200 \ \AA$.

\subsection{Comparison with STIS observations of H$\alpha$}
\label{sec_stis}

HST narrow slit spectroscopy of Ly$\alpha$ and $\Ha$ between 1999 -- 2004
has been discussed extensively by \cite{Michael2003} and
\cite{ Heng2006}, and recent observations from 2010 have been
discussed by \cite{France2010}. These papers were focused on the reverse shock properties, while we here emphasize the core component and the connection to our ground-based observations. Although the
spectral resolution of the G750L grating is only $\sim 450$ \kms
\ (FWHM), which is more than one order of magnitude less than UVES, it
is adequate for the broad lines. The most important advantage of the STIS data is
that it is possible to directly relate different spatial regions to
their velocities.
In the following discussion we will focus on the observations from 1999 and 2004,
where the slits covered
the full central region of the ejecta. The 2010 observation only covered the central $0.2$\arcsec \ and is therefore difficult to compare with the other observations.

\begin{figure}[!h]
\resizebox{\hsize}{!}{\includegraphics[width=1\linewidth]{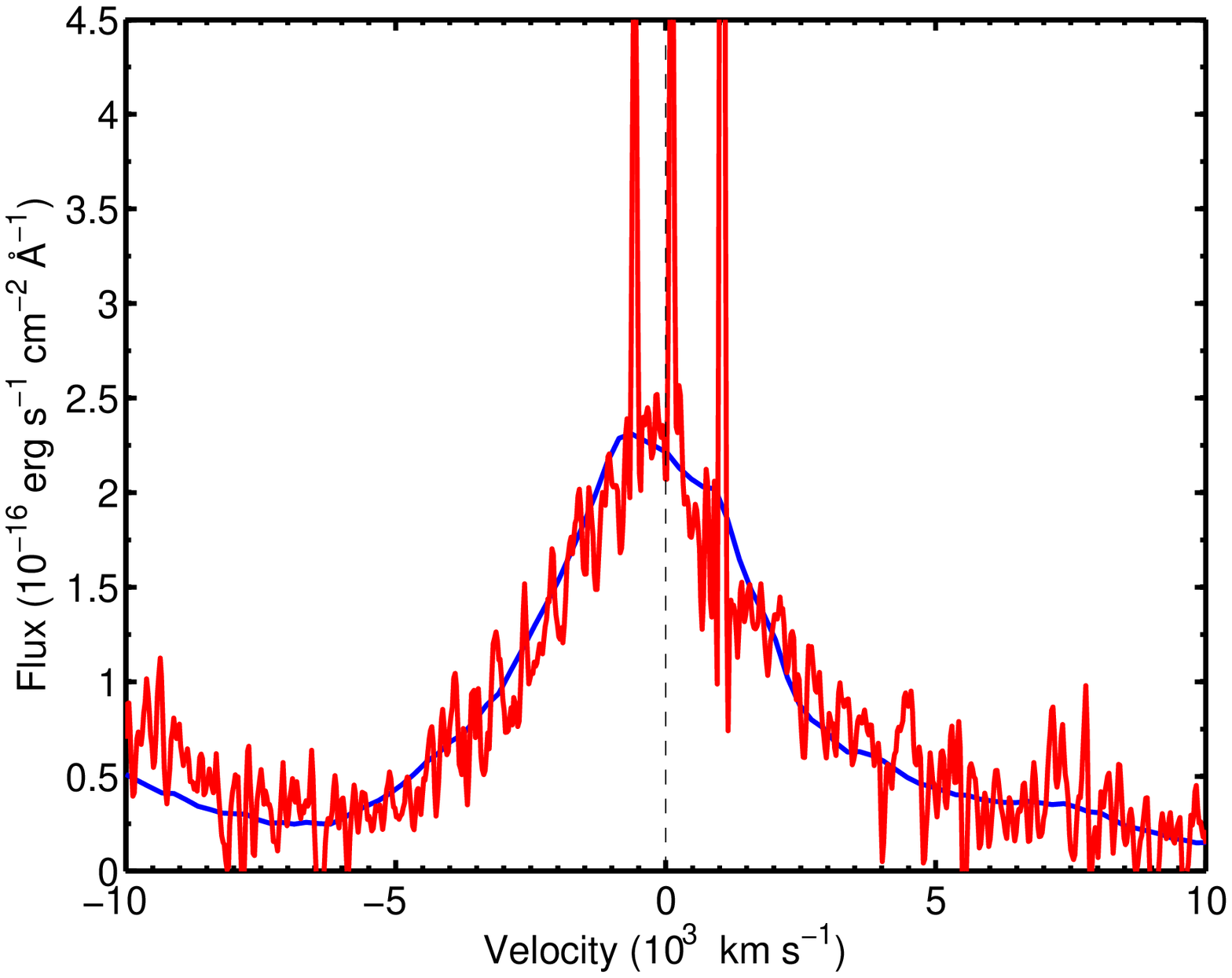}}
\resizebox{\hsize}{!}{\includegraphics{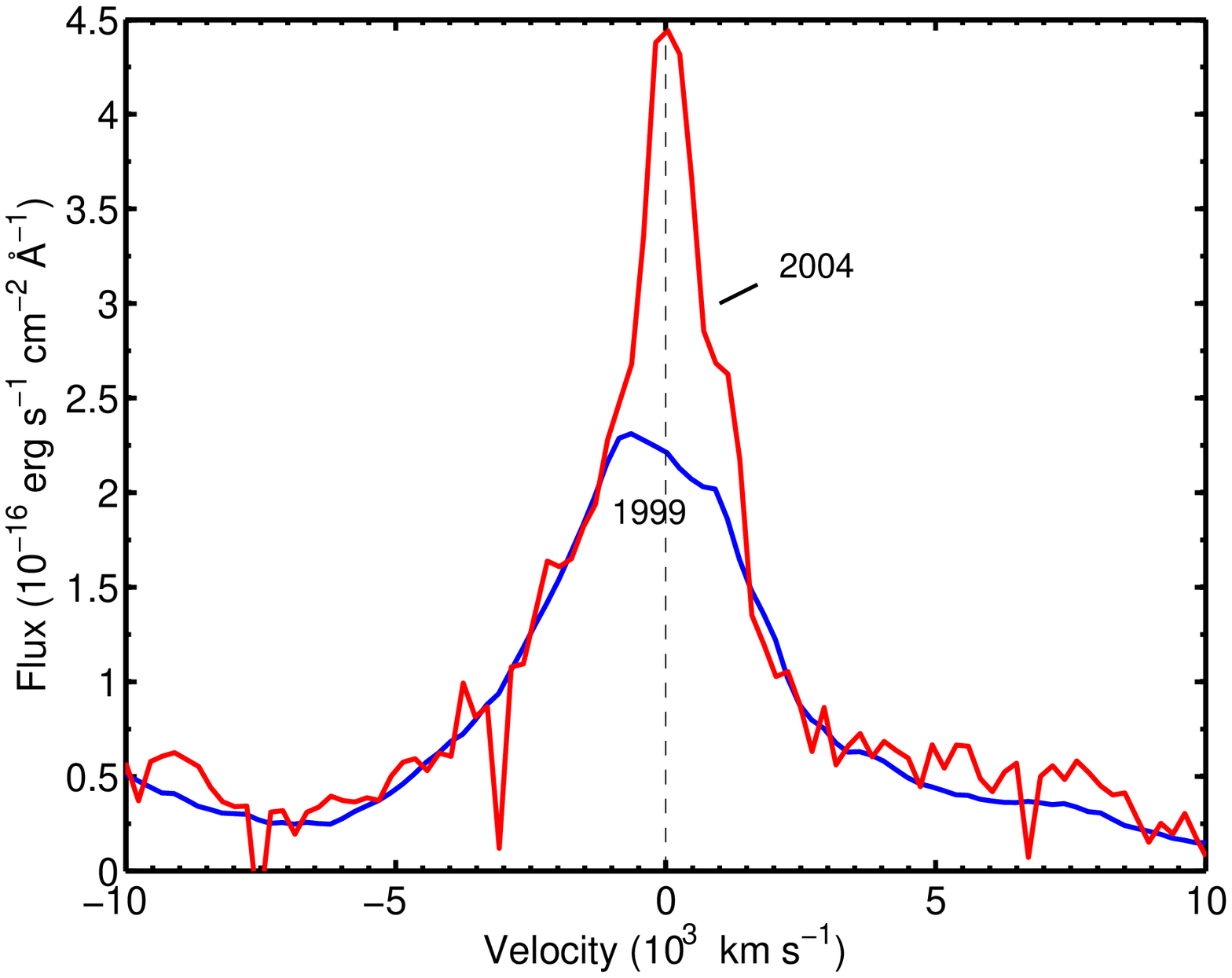}}
\caption{Upper panel: The STIS G750L (blue) and G750M (red) $\Ha$ line profiles from 1999 Feb.  (day 4381) and Aug. 1999 (day 4571-4572).  The G750M spectrum has been multiplied by a factor 1.5 to compensate for the smaller total slit width ($3 \times 0.1$ \arcsec) compared to the G750L spectrum ($0.5$ \arcsec). Lower panel:  STIS G750L $\Ha$ line profiles from  1999 Feb. (day 4381) and  2004 Jul. (days 6355-6360). The 2004 spectrum has here been multiplied by a factor 1.6 for comparison with the 1999 spectrum. The region between $\pm 1500 \ \kms$ in the 2004 spectrum is dominated by scattered $\Ha$, [N II] \wll 6548.1, 6583.5 emission from the ring collision.}
\label{fig_ha_1999_2004}
\end{figure}

The last  ejecta spectra not strongly affected by the ring collision are the STIS G750L and G750M spectra from 1999. The full G750L spectrum from  1999 Feb. (day 4381) is shown  in Fig. \ref{fig_ejecta_spectra}. This spectrum was taken with a 0.5\arcsec \ slit, which covered the full inner ejecta at this epoch. The spectra taken with the G750M grating (days 4571-4572) only cover the $\Ha$ line, but offer a much better resolution of  $\sim 50 \  \kms$. The observations were performed with three  0.1\arcsec \ slits, which together cover most of the inner ejecta. In the upper panel of Fig. \ref{fig_ha_1999_2004} we show a comparison of  the $\Ha$ profiles from the G750L and G750M gratings. Both spectra were extracted from the central $\pm 0.3$\arcsec \ along the slits. There is clearly a good agreement between the two spectra, with the only notable difference being that the medium resolution profile contains narrow lines from $\Ha$, [N II] \wll 6548, 6584 due to contamination from the outer ring, passing through the ejecta. We will discuss the line profile more below. 

\begin{figure*}
\centering
\includegraphics[width=17cm]{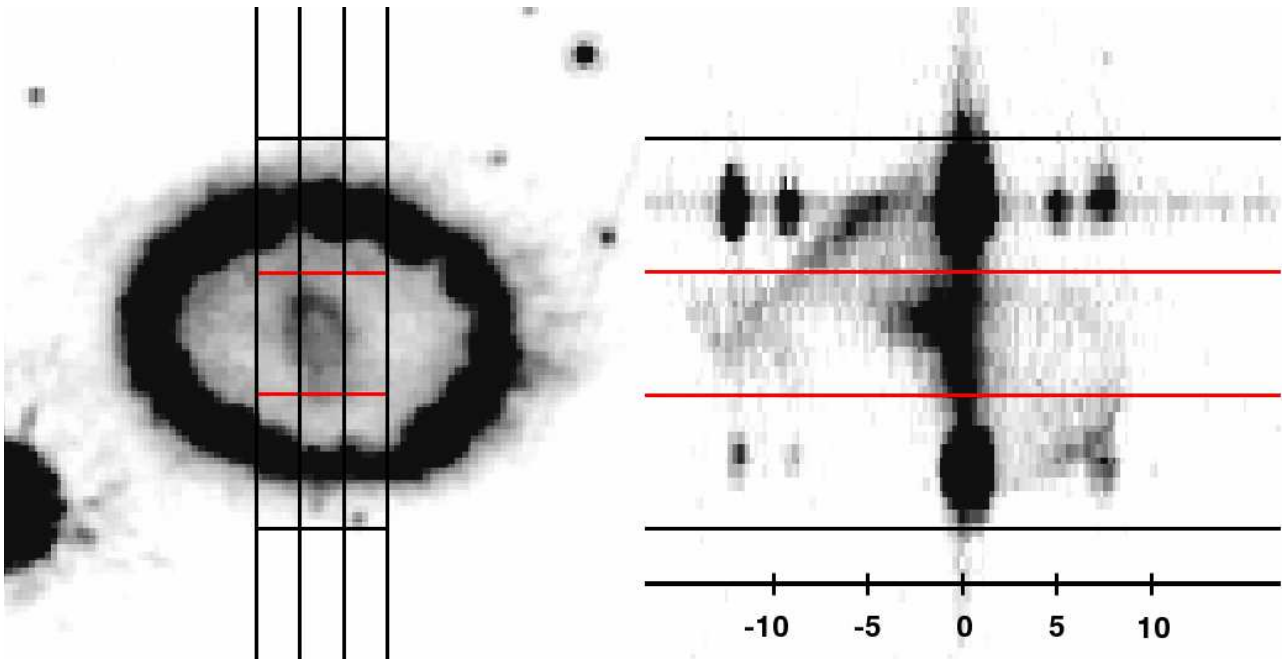}
\caption{Left panel: The HST/ACS F625W image from 2003 Nov. 28 (day 6122). The
  vertical black lines show the three slits positions used for the
   2004  Jul. (days 6355-6360) STIS observations. The slits are $0.2$\arcsec \
  wide. North is up and west is right. Right panel: The spectrum of the central slit from the STIS
  observations, taken with the G750L grating. The spectrum is centered
  on zero velocity for H$\alpha$. The horizontal red and black
  lines in both panels show the regions that were used to produce the
  spectra in Fig.~\ref{fig3d}. The spectrum in the right panel has been
  stretched in the y-direction to match the spatial scale of the image
  on the left. The velocity scale is in $1000 ~ \kms$ \ from the rest wavelength of H$\alpha$.}
\label{fig3f}
\end{figure*}

In the later 2004 spectrum, the ring interaction is much more prominent. Figure \ref{fig3f} shows an ACS image of SN~1987A taken in the F625W
filter on 2003 Nov. 28  (day 6122)  with the three slit positions, each 0.2\arcsec
\ wide, as well as the 2-dimensional spectrum from the central slit. The latter is
similar to Fig. 4 in \cite{Heng2006}. We
include it here to illustrate the area of extraction for the spectra.

The full STIS spectrum from the 
central area indicated in Fig.~\ref{fig3f}  is shown in Fig. \ref{fig_ejecta_spectra}. The degree of contamination from the outer ring and scattered light in the central part of the line is difficult to estimate. We plot the 2004 spectrum together with the 1999 low resolution spectrum in the lower panel of Fig. \ref{fig_ha_1999_2004}. The 2004 spectrum is the total over all three slits, scaled to the same level as the 1999 spectrum outside of $\pm 2000 \ \kms$.  We first note the similar shape of the wings between $2000 - 5000 \  \kms$ on both the red and blue sides. There is little evolution between these epochs in these velocity intervals. The central part of the 2004 spectrum is dominated by scattered light from the shocked ring emission as evidenced by the fact that the $\Ha$/[N II] ratio is similar to that of the shocked ring \citep{Groningsson2008b}.

\begin{figure}[!h]
\resizebox{\hsize}{!}{\includegraphics[width=1\linewidth]{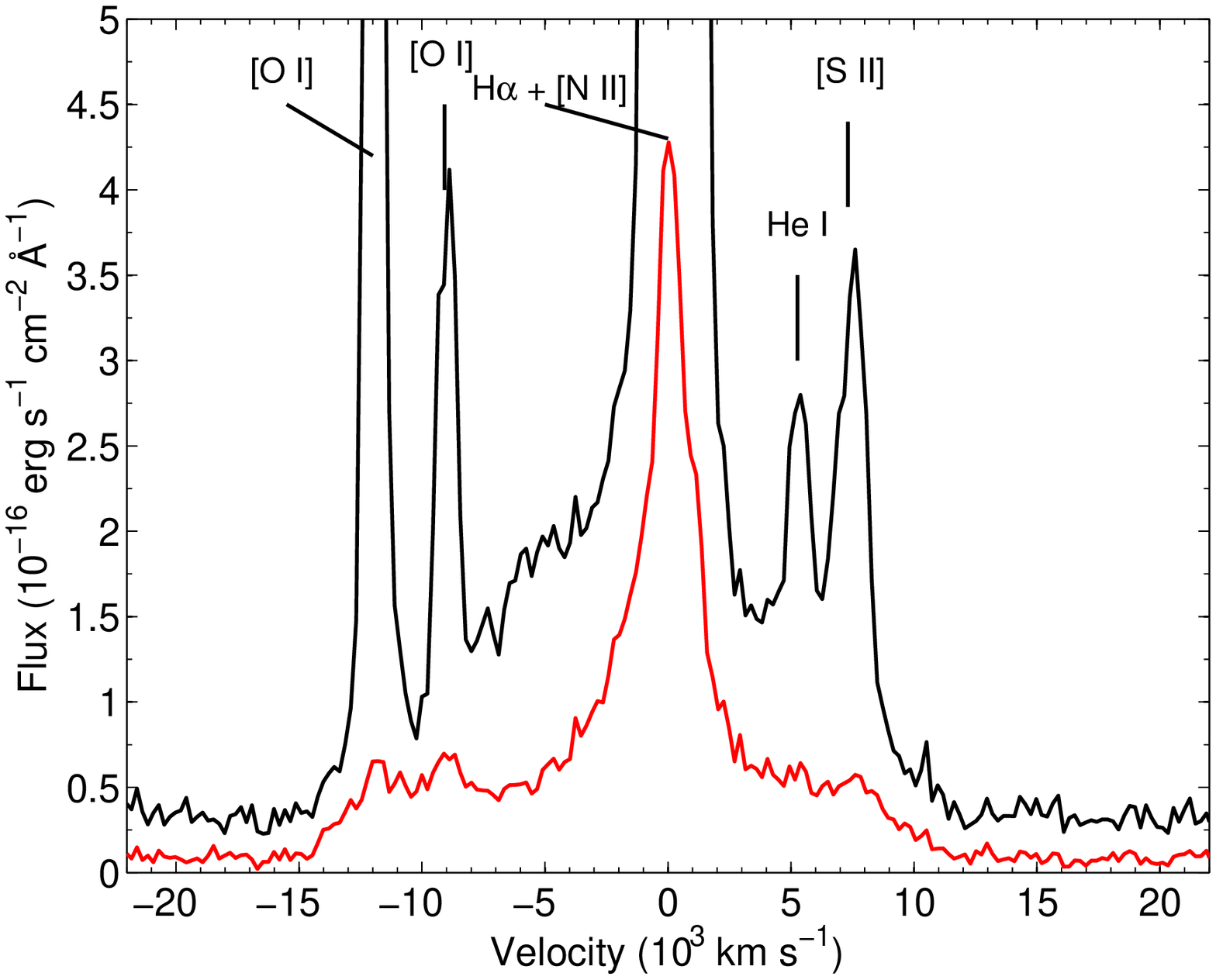}}
\caption{STIS G750L spectra from  2004 Jul.  18-23  (day 6355-6360). The spectra were
  extracted from the areas shown in Fig. \ref{fig3f}, and are shown
  with the same color coding as in that figure. The red, bottom spectrum was extracted from
  the central $\pm 0.3$\arcsec \ and the black, top spectrum is from $\pm
  0.9$\arcsec. 
 }
\label{fig3d}
\end{figure}

The 2-dimensional STIS spectra clearly display a high
velocity component close to the ring on both the red and blue
sides. This is the freely expanding, neutral reverse shock
component. Most of this emission comes from $\sim 3000 - 8000$ \kms
\ on the blue side and from $\sim 4000 - 7000$ \kms \ on the red
side. The blue component is fairly uniformly distributed over the
three slit positions close to the northern part of the ring, while the
fainter component on the red side mainly comes from the south-eastern
side of the ring. \cite{ Heng2006} find that the flux from the reverse shock component in $\Ha$
increased by a factor $2 - 3$ from 1999 Sep. 18 to 2004 Jul. 18 (days 4589 and 6355, respectively).

The increase in the northern component seen by \cite{Heng2006} can
also be seen in our UVES and FORS spectra in Fig. \ref{fig4} on the
blue side at $\sim -5000$ \kms. 
A similar increase on the red side from the southern
part of the reverse shock can also be seen. The high spatial
resolution observations by HST are therefore consistent with our VLT observations.

We extracted one-dimensional spectra from the regions marked by red and black lines in Fig.~\ref{fig3f} and added the resulting spectra from all three slit positions. Figure \ref{fig3d} shows the
resulting $\Ha$ line profiles with different sampling in the N -- S
direction from the center. The similarity of the $\Ha$ line from the
full area (area within the black lines in Fig. \ref{fig3f}) with the UVES
 2005 Mar. spectrum in Fig. \ref{fig4} is noteworthy. Although the S/N, as well as
the spectral resolution, of the STIS spectrum is much worse than the
UVES spectrum, one can recognize several similarities. In particular,
the flat regions at 3000 -- 8000 \kms \ on both the red and blue sides
are apparent in this integration. The low velocity part of the line is completely dominated by the narrow lines from the ring collision. 

 In the next extraction  from the inner
region between -0.3\arcsec \ and +0.3\arcsec \ (red lines) the high velocity component of the line decreases by a
factor $\ga 3$. This is clear evidence that
this spectral feature originates in the reverse shock, appearing as a thin
streak in Fig. \ref{fig3f} (right panel). The central  region  is  dominated by the ejecta component.
There is, however, still a flat region at both high positive and negative
velocities, originating from the reverse shock in the line of sight  towards the center of the ejecta.

The spatial resolution of STIS allows us to isolate the core
component, and in this way check the VLT observations.  Assuming that the emission from the core has the same profile as in 1999 (Fig. \ref{fig_ha_1999_2004}) we find for the $\Ha$ flux on day 6355  that the flux is nearly a
factor of two larger than the flux from the UVES measurement that is
closest in time (from day 6622)  (Fig.  \ref{fig_core_rev}). A comparison between Fig. \ref{fig4}
and the red spectrum in Fig. \ref{fig3d} shows that there are two main
reasons for this. Firstly, the blue wing of the UVES core component is
lost due to the interpolation between $-1000$ and $2000$ \kms and,
secondly, some of the UVES low-velocity component (especially on the
red side) has been lost in the process of removing the narrow
lines. 

\begin{figure}[!h]
\resizebox{\hsize}{!}{\includegraphics[width=1\linewidth]{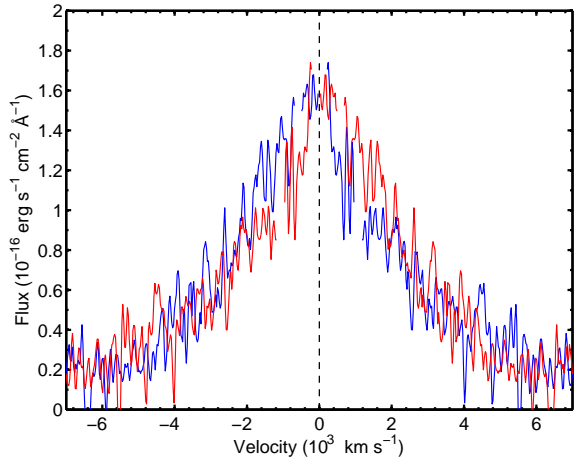}}
\caption{Comparison of the red and blue wings of $\Ha$ from the STIS G750M 1999 spectrum. The blue spectrum is the original spectrum, while the red has been reflected around zero velocity.  The narrow $\Ha$ and  [N II] \wll 6548, 6584 lines seen in Fig. \ref{fig_ha_1999_2004} have been removed . }
\label{fig_ha_symm}
\end{figure}

The line profile from the low resolution STIS spectra of the ejecta shows some indication of an asymmetric
velocity line profile, with more emission on the blue side. Unfortunately, the low velocity part of the line profiles in the G750L spectra is severely affected by scattered ring emission. We therefore concentrate on the medium resolution spectrum from 1999. To study the asymmetry of the line we have reflected the line profile around zero velocity, and compared this with the original in Fig. \ref{fig_ha_symm}. 

When we compare the red and blue line profiles we see that for $\ga 2000 \kms$ the blue and red wings are nearly identical. At lower velocities the blue wing is $\sim 15 \%$ stronger. We therefore find that solely on the integrated line profile  there is only weak evidence for an asymmetry from the $\Ha$ line. 

The 2D spectra in Fig. \ref{fig3f}, on the other hand, show strong blue emission from especially the northern side, while there is little emission from the southern side. This originates in the inner ejecta and extends
from zero velocity to $\sim -5000$ \kms.  This red/blue asymmetry is consistent with dust
absorption of the emission from the far side of the
ejecta. 

The message we get from the integrated line profiles and the 2D spectra may seem somewhat contradictory. It does, however, show that even strong asymmetries may cancel in the integrated spectra and ideally one needs the full 2D (or even better 3D) information. This is discussed in more detail in L13.

\subsection{Metal lines from the inner core}

\begin{figure}[!h]
\resizebox{\hsize}{!}{\includegraphics{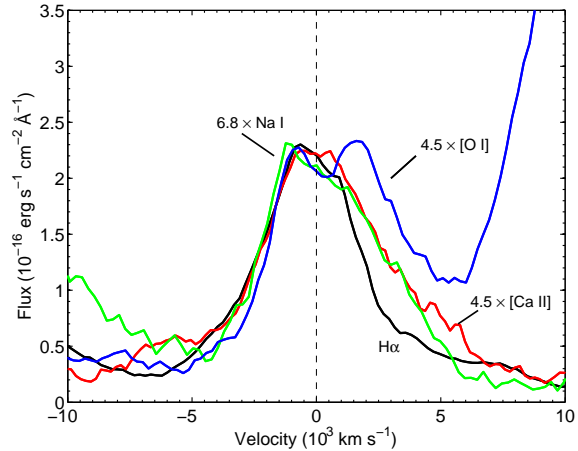}}
\caption{Comparison of the H$\alpha$, Na I \wll 5890, 5896, [O I] \wll 6300, 6364, and [Ca II] \wll
7292, 7324 
from the STIS G750L observation from  1999 Feb. (day 4384). For a direct comparison of the different profiles the fluxes of the lines have been scaled by a factor given in the figure. The doublet lines have been centered on the blue components.}
\label{fig_stis_profiles}
\end{figure}

Although not covering the blue part, the STIS 1999 (day 4384) spectrum in Fig.  \ref{fig_ejecta_spectra} shows the most important metal lines, as well as H$\alpha$. The distribution of the different elements can be inferred from
their line profiles. In Fig. \ref{fig_stis_profiles} we show these, centered on the blue doublet components. When we compare the blue wings of these lines we see that the Na I and [Ca II] lines have very similar profiles to H$\alpha$, while the [O I] \wl 6300 line has a smaller blue extent. This is consistent with the results in \cite{Jerkstrand2011} where the Na I and [Ca II] lines are dominated by emission in the H envelope, while the [O I] lines arise mainly in the O rich core. 

Note that the [O I] \wl 6364 line sits on top of the flat reverse shock component from the H$\alpha$ line (Fig. \ref{fig_ha_1999_2004}), which partly explains its comparable strength to the [O I] \wl 6300 component (usually $1/3$ of the latter).  In addition, there may be a contribution to these lines from Fe I \wll 6280, 6350 \citep[see Fig. 4 in][]{Jerkstrand2011}.

Except for H$\alpha$, easily identified broad lines in the UVES spectra are [Ca II] \wll
7292, 7324 and Mg I] \wl 4571.  There is  also a clear feature coinciding with the [O I] \wll 6300, 6364 doublet. In addition, Fig.  \ref{fig_ejecta_spectra} shows an increasingly strong line feature at $\sim 9210$ \AA. In Sect. \ref{sec_FeII} we identify this with Mg II  \wll
9218, 9244.  Of these the [Ca II] lines, and at the last epochs the Mg II lines, are both the strongest and best defined.

\begin{figure}[!h]
\resizebox{\hsize}{!}{\includegraphics{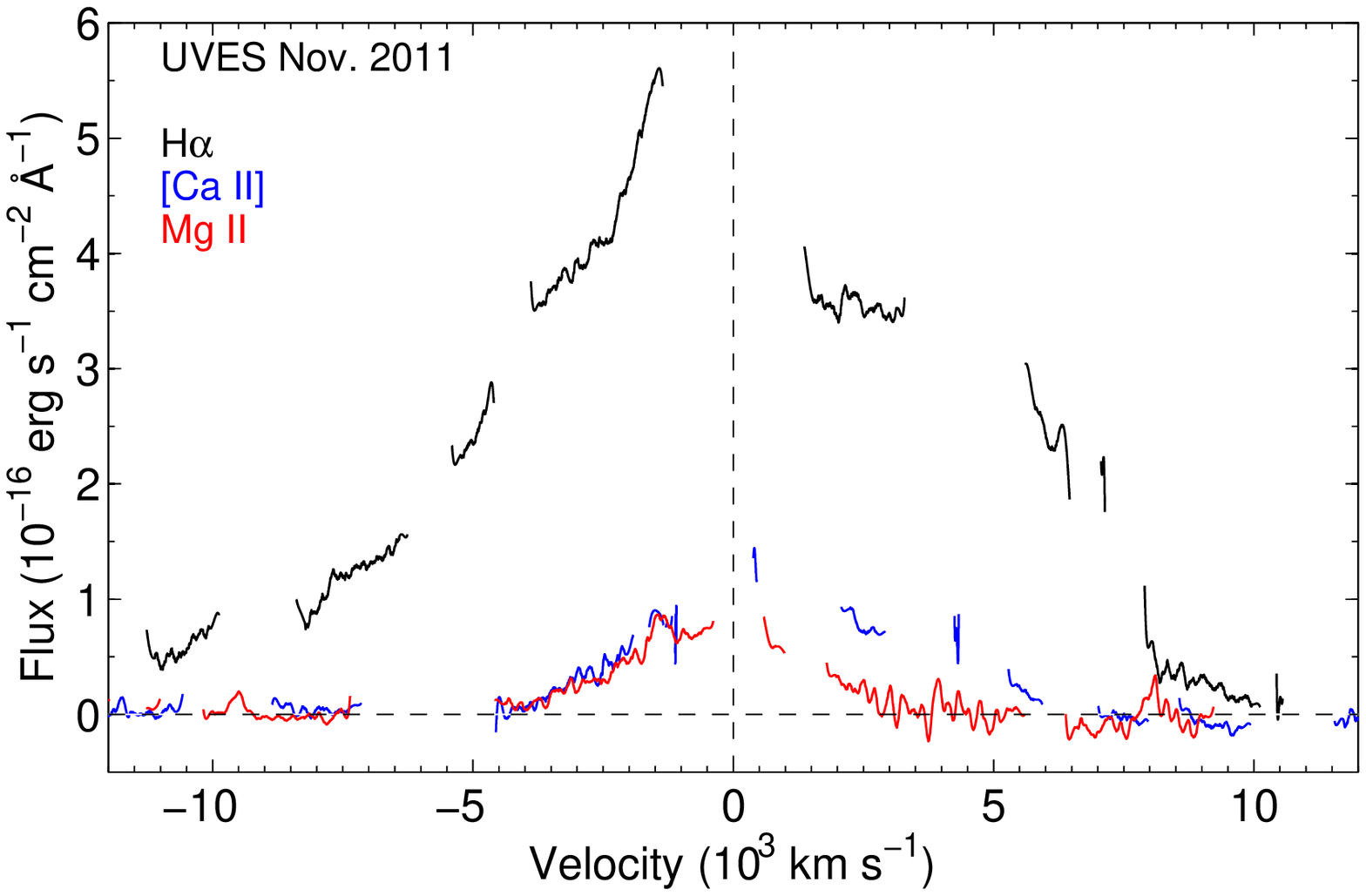}}
\resizebox{\hsize}{!}{\includegraphics[width=1\linewidth]{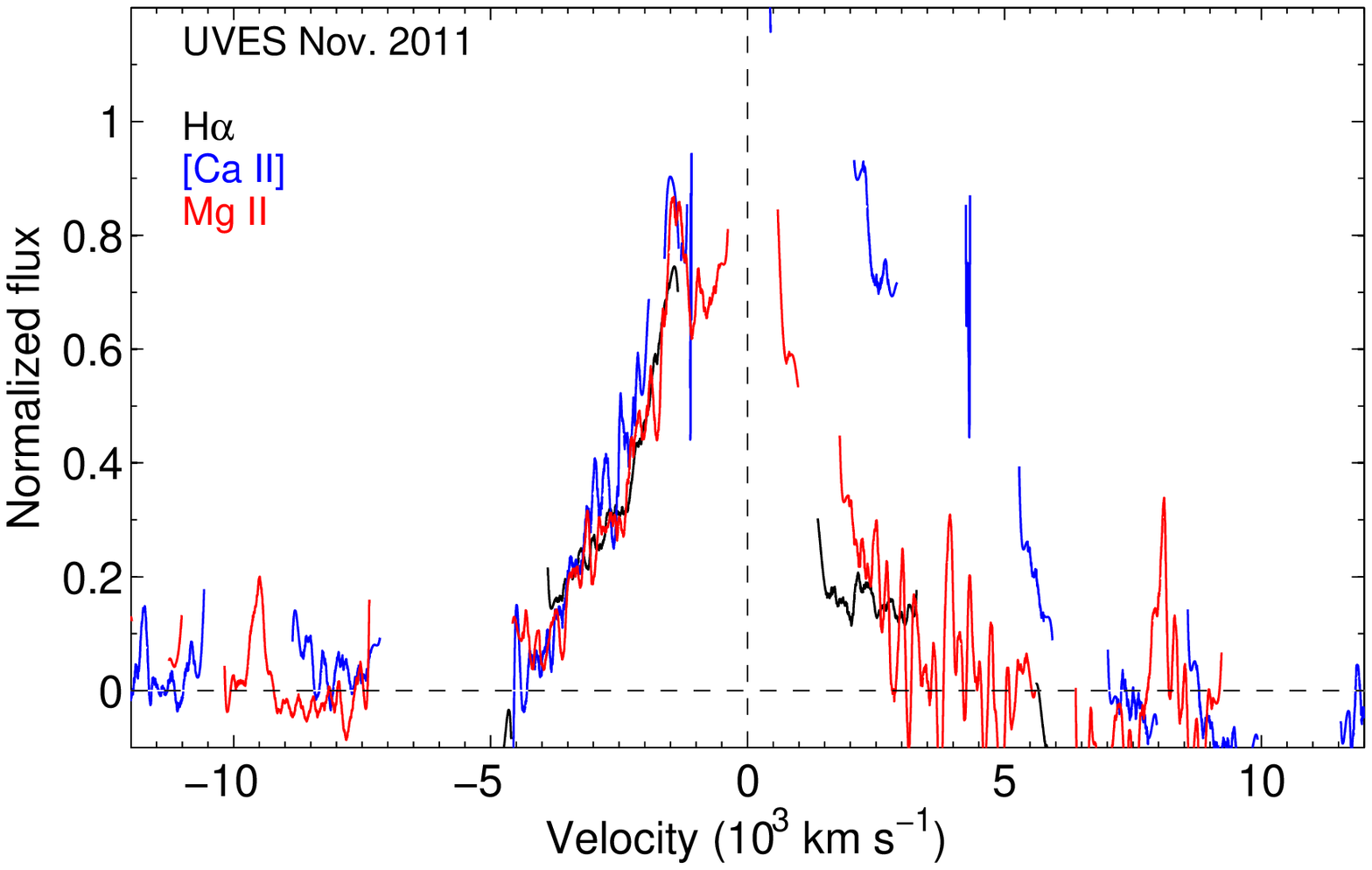}}
\caption{Upper panel: Comparison of the H$\alpha$, [Ca II] and Mg II  \wll
9218, 9244
profiles at 9019 - 9035 days. Lower panel: Normalized line profiles of the same lines. For H$\alpha$ the 'continuum' level has been set at  $\sim 3.5 \times 10^{-16} \ \ergs \  {\rm cm}^{-2}  \ \AA^{-1}$ in the upper panel. Note the similar
profiles of the blue wings of the [Ca II] and Mg II lines to the core of $\Ha$. }
\label{fig6}
\end{figure}

As mentioned in the introduction, and discussed in Section \ref{sect_xrays}, the energy
input after $\sim 5000$  days comes from X-rays produced by the collision with the
ring more than from radioactivity. 
To compare the line profiles at the latest stages when X-rays dominate, we show in Fig. \ref{fig6} H$\alpha$, [Ca II] \wll 7292, 7324  and the  Mg II  \wll 9218, 9244 lines from day 9019 - 9935. We have here removed the narrow lines, as before, and also subtracted a continuum level from each line. To more easily compare the core component of the lines we show in the lower panel the same lines normalized to the same level for the blue wing, which is well defined for all three lines.  To isolate the core component we have for H$\alpha$ put the 'continuum' level close to the 'shoulder' on the red wing (at $\sim 3.5 \times 10^{-16} \ \ergs \  {\rm cm}^{-2}  \ \AA^{-1}$ in the upper panel).

When comparing the [Ca II] and
  $\Ha$ lines in Fig. \ref{fig6} the former lacks the flat, boxy part
  of the line profile. As the lower panel shows, the central part of $\Ha$ and the blue wing of the two other lines, below $\sim
  4000 - 5000$ \kms \ are, however, very similar and argue for a similar
  origin.  The full extent on the blue side is difficult to
  estimate because of contaminating lines from the ring, but reaches at least 4000 \kms. On the red
  side the lines are less well-defined, and also affected by the blue components of the [Ca II] and Mg II doublets.   The Mg I] \wl 4571 line
    has a lower S/N, but its profile is consistent with that of the
    [Ca II] line.

\begin{figure}[!h]
\resizebox{\hsize}{!}{\includegraphics[width=1\linewidth]{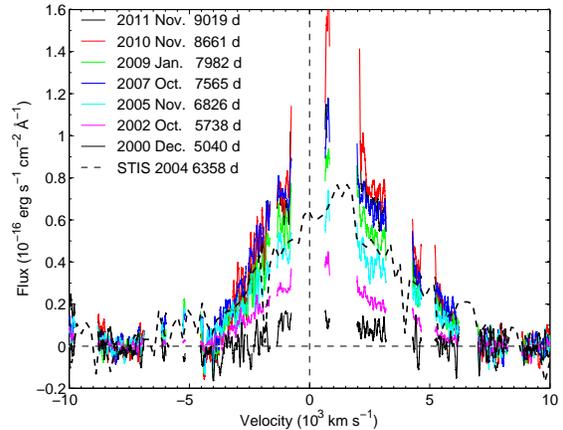}}
\caption{ Same as Fig. \ref{fig4} for the [Ca II] \wll 7292, 7324
  lines. Velocities are measured from the \wl 7292 component. For comparison we have also added the STIS spectrum for day 6358 as the dashed line.}
\label{fig6b}
\end{figure}

Similarly to the other broad lines the line profile of the [Ca II] line is
contaminated by several strong lines from the ring collision. We can
nevertheless trace most of it by interpolation
(Fig. \ref{fig6b}). This provides a good representation of the
 true line profile as confirmed by the comparison of the [Ca
  II] line profiles from the Jul. 2004 STIS spectrum and the 
2004 Apr. UVES spectrum, shown as the dashed line in the figure.

\begin{figure}[!h]
\resizebox{\hsize}{!}{\includegraphics[width=15cm,angle=-0]{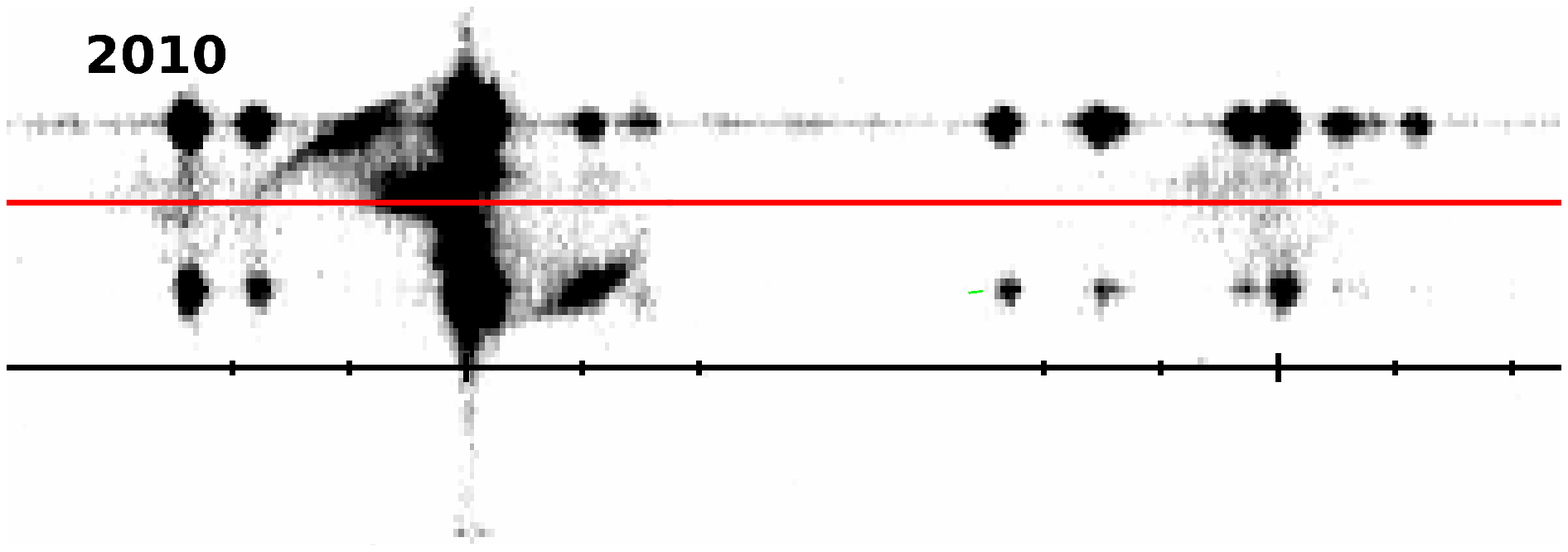}}
\resizebox{\hsize}{!}{\includegraphics[width=14cm,angle=-0]{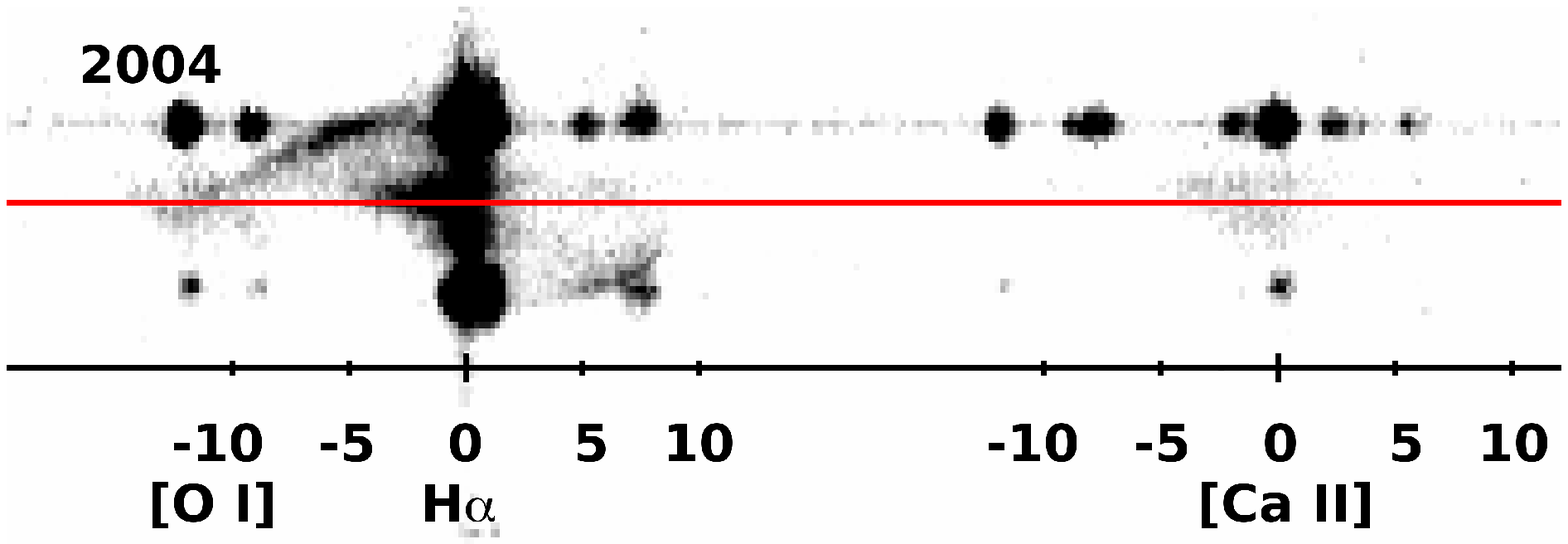}}
\caption{Two dimensional STIS spectra from 2004 (lower) and 2010 (upper) of the $\Ha$ (left strong line)
  and [Ca II] (right fainter line) regions from the central slit in
  Fig. \ref{fig3f}. To show the faint [Ca II] line better the
  intensity levels have been shifted compared to
  Fig. \ref{fig3f}. Note the absence of any high velocity reverse
  shock component similar to that of $\Ha$ for [Ca II]. We also show the position of the [O I] doublet, which can be seen also in the ejecta with a velocity distribution similar to the [Ca II] lines. The velocity scales are in $1000$ \kms \  from $\Ha$ and  [Ca II], respectively. }
\label{fig6bb}
\end{figure}

Figure \ref{fig6b} shows
that the general line profile remains roughly the same over the whole
period. In particular, the maximum velocities of the red and blue
wings remain at $\sim \pm 5000$ \kms. This and the very different
shape of the line in comparison with $\Ha$ argue strongly for this line originating 
in the core and not from the reverse shock region. This is also
confirmed by the spatial location in the STIS spectrum of the [Ca II]
region (Fig. \ref{fig6bb}) which shows a central component extending to $\sim
5000$ \kms.

The absence of a reverse shock component in the [Ca II] lines (as well
as other ionized species) is mainly a result of the low abundance in the envelope.  The combination of a forbidden transition and a low ionization potential also means that the number of excitations per ionization will be low. This is in contrast to the Li-like ions, like C IV, N V and O VI, which have resonance lines far below the ionization level. This results in a large number of excitations for each ionization \citep{Laming1996}.  This
compensates for the low abundance and produces lines that are as strong as
those of H and He. This is, however, not the case for the [Ca II] and the other forbidden lines in the optical, like the [O I] and Mg I] lines. We also note that because of the positive charge the Ca II ions will, in contrast to H I, be affected by the magnetic field in the shock. Their velocity will therefore quickly be isotropised and the velocity broadening will therefore be characteristic of the post-shock velocity.

\begin{figure}[!h]
\resizebox{\hsize}{!}{\includegraphics{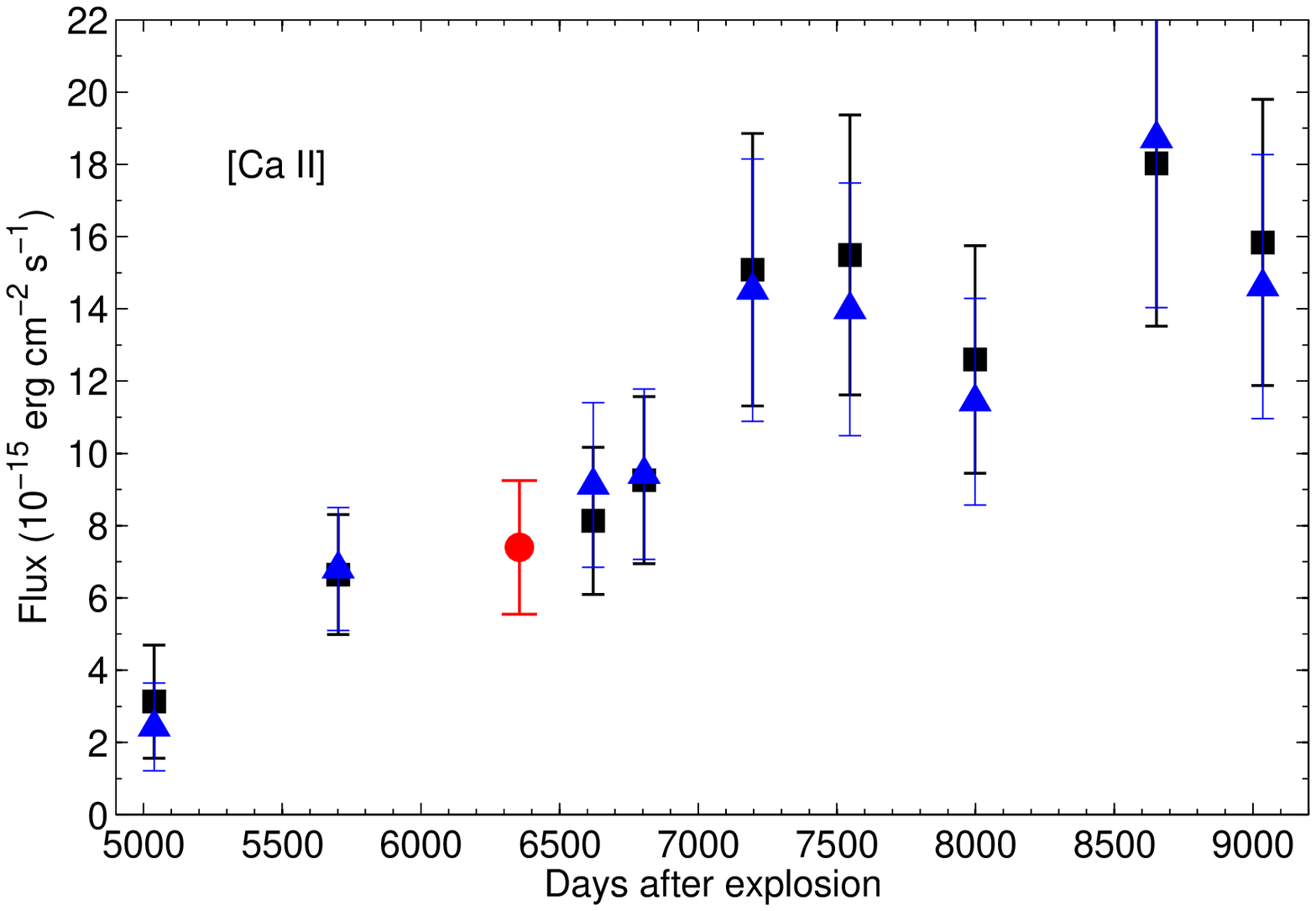}}
\resizebox{\hsize}{!}{\includegraphics{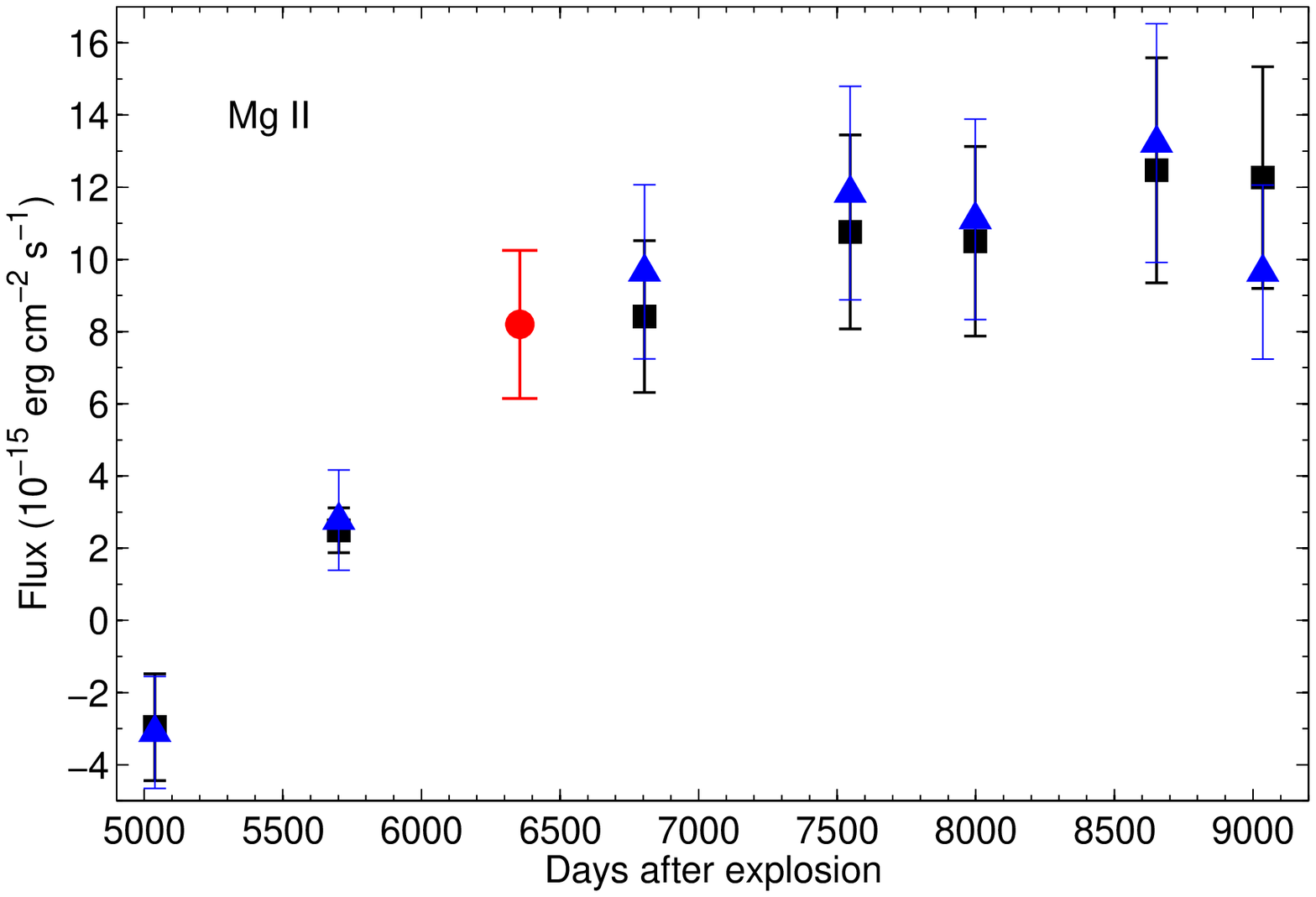}}
\caption{Upper panel: Evolution of the [Ca II] \wll 7292, 7324 flux with time with UVES and STIS. The squares are from UVES and the circle point at 6355 days from STIS.  The triangles show the flux from the regions of the line not affected by the narrow lines, multiplied by a factor 2.5. Lower panel: Same for the Mg II  \wll
9218, 9244. The fluxes from the unblocked regions of the lines have in this case ben scaled by a factor 1.4. }
\label{fig7}
\end{figure}

The upper panel of Fig. \ref{fig7} shows the evolution of the total flux of the [Ca II]
lines from the UVES observations. The squares show the flux determined from the interpolated line profile where the missing sections of the line in Fig. \ref{fig6b} have been replaced by spline interpolations. To estimate the uncertainty introduced by this interpolation we have also calculated the flux from the line, omitting these regions. These are shown as triangles. For a comparison with the interpolated line we have multiplied these fluxes by a constant factor 2.5. As an extra check we also show the flux
from the STIS 2004 observation measured from the central area shown in
Fig. \ref{fig3f}. This flux ($7.4 \times
10^{-15}\ \rm{erg\ cm^{-2}\ s^{-1}}$ on day 6355 after the explosion)
is similar to that found from UVES, and well within the
error bars. 

We see in Fig. \ref{fig7} that the flux increased monotonically by a factor of $\sim 4-6$ up to day $\sim$ 7000. At later epochs it is nearly constant. This increase in flux
is an important indication that there is additional energy input to
the inner ejecta in addition to the radioactive energy source. In L11 these lines were used as an independent confirmation of the HST photometry, where the spatial information was combined with the photometry to show that the increased flux was coming from the inner ejecta, and not from the reverse shock region.  

At these stages, the [Ca II] \wll 7292, 7324 lines are mainly excited by fluorescence of UV
emission through the H \& K lines \citep{Li1993, Kozma1998II, Jerkstrand2011}. Therefore the Ca II triplet \wll
8498, 8542, 8662 is expected to have the same total flux as the
\wll 7292, 7324 lines, unless collisional de-excitation is important. The STIS spectrum in  Fig. \ref{fig_ejecta_spectra} shows that
there indeed is a  line feature at this wavelength range,
and with a flux consistent with that of the \wll 7292, 7324 lines, although it is more smeared out in wavelength than the \wll 7292, 7324 lines.

We also note the presence of an additional broad line with a peak wavelength close to that
of He I \wl 5876/Na I \wll 5890, 5896. \cite{Chugai1997} claim that this is Na I and based on
modeling by \cite{Jerkstrand2011} this is a likely identification.

\subsection{The \wl 9220 \AA \ feature}
\label{sec_FeII}
The broad line at $\sim$ 9220 \AA \  has no obvious identification.  The full extent is $\pm 3500$ \kms,
indicating an origin in the core of the supernova. It is unlikely that this
is a blend of lines from the shocked ring, which have a width of $\pm
300$ \kms, since the individual lines would then be resolved. This is
also confirmed from a direct inspection of the STIS spectra, which
shows that the \wl 9220 feature has the same spatial distribution as the
[Ca II] \wll 7292, 7324 lines.  Possible identifications include
the Paschen 9 line at 9229.0 \AA. In this case we would, however, expect to see
P10 at 9015 \AA\ nearly as strong and P8 9546 a factor $\sim 1.5$
stronger. Neither of these are seen at this level. In addition, for
pure Case B recombination this line is expected to have a flux of
$\sim 3 \%$ of $\Hb$ \citep{Hummer1987}, which is too weak to explain
the line. 

Another possible identification is a blend of Fe II lines, powered
by Ly$\alpha$ fluorescence.  This is the mechanism which produces the Fe II lines in SN 1995N , but in that
case, the Ly$\alpha$  was produced in the circumstellar interaction  \citep{Fransson2002}..

\cite{Sigut1998,Sigut2003} have calculated Fe II spectra including
Ly$\alpha$ excitation from the first excited state, a${}^4$D, to
higher levels, in particular u${}^4$P, u${}^4$D and v${}^4$F. The
cascade from these to lower levels result in a number of strong Fe II
lines, with the strongest being the 9122.9, 9132.4, 9175.9, 9178.1,
9196.9, 9203.1 and 9204.6. 

Observationally, \cite{Hamann1994} see a strong feature at 9216.5 in the spectrum
of Eta Carinae. They identify this mainly with Fe II \wll 9203,
9204. Based on their observation, also Fe II 8490.1 is then expected
to be strong. These authors claim all these lines to be pumped by
Ly$\alpha$.  A weaker broad feature at this position is probably
present in the STIS spectrum in Fig. \ref{fig_ejecta_spectra}, so this may be
consistent. The same feature is also seen in the symbiotic
nova PU Vul by \cite{Rudy1999} and again identified as Fe II \wll 9176--9205.

For the Ly$\alpha$ pumping mechanism to work the Ly$\alpha$ flux needs
to be strong and the population of the Fe II a${}^4$D level large
enough for fluorescence to be efficient. In addition, the width of
the Ly$\alpha$ line should be large enough to have a substantial flux at the Fe II line. The first and third
requirements are certainly fulfilled. Both the Ly$\alpha$ flux from the
ring collision \citep[e.g.,][and references therein]{Heng2006} and
from the radioactive excitation of the ejecta \citep{Jerkstrand2011}
are strong. A major problem is, however, to understand the excitation
to the a${}^4$D level from which the fluorescence takes place.  The excitation energy corresponds to $\sim
11,000$ K, which is much larger than the temperature expected in the radioactively
powered core, 100 -- 200 K \citep{Jerkstrand2011}.  The level is
powered by non-thermal excitations and recombinations, but these are
unlikely to be sufficient to make the line optically thick. We
therefore conclude that the Ly$\alpha$/ Fe II fluorescence is unlikely to
work here.

A more plausible identification of the \wl 9220 feature is Mg II \wll
9218, 9244 from the 4p ${}^2$P${}^o$ level. This line can
arise either as a result of fluorescence by Ly$\alpha$ or
Ly$\beta$. The 5p ${}^2$P${}^o$ level is connected to the Mg
II ground state by \wll 1026.0, 1026.1, in nearly perfect resonance
with Ly$\beta$ at 1025.7 \AA. This decays to 4p ${}^2$P${}^o$
via either the 5s ${}^2$S or 4d ${}^2$D level, with
emission at 8214, 8235 and 7877, 7896, respectively. This has
earlier been suggested to explain UV and IR lines seen in QSOs and
Seyferts by \cite{Grandi1978} and \cite{Morris1989}. In our case this
has a problem in that the \wll 7877, 7896 and \wll 8214, 8234
lines are expected to have similar strengths as the \wll 9218,
9244 lines, but are not seen in the spectra. The efficiency of the
Ly$\beta$ / Mg II fluorescence compared to branching into $\Ha$ is also
expected to be low.

Instead, we propose that the lines arise as a result of Ly$\alpha$
fluorescence directly to the 4p ${}^2$P${}^o$ level. The wavelength
differences to the multiplet levels are 1239.9, 1240.4 \AA. A velocity
shift of $\sim 6000$ \kms \ will therefore redshift the Ly$\alpha$
photons into these lines, unless other optically thick lines are
present at shorter wavelength. A velocity difference of this magnitude
can e.g., arise from a photon emitted on the far side of the core and
absorbed on the near side,

In the analysis of the 8 year spectrum in \cite{Jerkstrand2011} it was indeed found that the Mg II line
dominates the production of the \wl 9220 emission. Magnesium is mainly in the
form of Mg II in the hydrogen rich regions, and the optical depth of
the \wll 1239.9, 1240.4 lines are larger than unity in the central
regions within the supernova core, i.e. within $3000-4000$ \kms. This remains true also at $\sim 20$ years and we therefore expect the fluorescence to be effective also at these epochs. We
also find that the lines at \wll 10,914 --10,952 should have a
strength a factor $\sim 2$ fainter than the 9220
feature. Inspecting one of the few existing NIR spectra of the ejecta in \cite{Kjaer2010} there is a strong feature at this wavelength, although this region is noisy and contaminated by emission from the ring. This therefore supports the identification above. Other lines  expected are the \wll 2929 ,
2937 lines in the UV.  These are, however, likely to be scattered by the many
optically thick resonance lines in the UV.

As is apparent from Fig. \ref{fig_ejecta_spectra}, the flux of this line increases rapidly with time. In Fig. \ref{fig7} we show the total line flux from the spline interpolated line profile as squares. As in the previous section, we also show the flux from the unblocked sections of the line as triangles. Because a smaller fraction of the line is affected by narrow lines, we only have to multiply this by a factor of 1.4 to bring it to the same level as the full, interpolated flux. 
The first point at 5038 days shows a negative flux due to a combination of the noise and level of the subtracted continuum.  It is
therefore an estimate of the errors in the flux determination.

When we compare it with the evolution of the [Ca II] lines in the upper panel we see the same basic evolution. The scatter of the points is, however, smaller due to the smaller fraction of line blocked by narrow lines.

\section{Discussion}
\label{sec_discuss}
The broad lines extracted from the spectra of SN 1987A in this paper
represent observations of a supernova ejecta in the
transition from radioactively powered supernova to one powered by the
interaction with the CS environment. Only a few supernovae dominated by
circumstellar interaction already after the first year, like SN 1979C \citep[][]{Milisavljevic2009}, can compete with
this. Compared to these we have in the case of SN 1987A also spatial
information, which is crucial in interpreting the observations.

The emission from the ejecta can clearly be separated into two
components. One of these comes from the inner regions of the supernova
ejecta, well represented by the [Ca II] \wll 7292, 7324 lines.
This is the same component as was seen in the spectrum at 5 -- 6 years and
7.8 years by \cite{Wang1996} and \cite{Chugai1997}, respectively. The
other component, seen in the Balmer lines, comes from the reverse
shock, resulting from the interaction with the circumstellar
ring. The Balmer lines have, however, also a lower velocity component from the inner parts of the ejecta. Because of the different origins of the two components we discuss them one by one in the next sections.  

\subsection{Reverse shock}
\label{sec_rev_shock}
The reverse shock in SN 1987A has been discussed by e.g.,
\cite{Michael1998a,Michael2003,Smith2005,Heng2006} and \cite{France2010}. The
most important results in this paper regarding the reverse shock are  
the evolution of the line profiles, as well as fluxes. 

As mentioned in Sect. \ref{sec_introd}, \cite{Smith2005} discuss the
evolution of the flux from the reverse shock extensively. A
particularly interesting point was their prediction that, around 2012--14, the H$\alpha$
flux from the reverse shock should reach a maximum and then gradually
vanish. 
The rationale is that the EUV and soft X-ray
flux from the shocks ionizes the neutral hydrogen before it reaches
the reverse shock. The protons are then accelerated in the shock
and do not emit any line radiation, except by charge transfer with any unshocked H I still present.

From the H$\alpha$ luminosity and the estimate that 0.2  H$\alpha$  photons are emitted for each hydrogen ionization \cite{Smith2005} estimate a flux of $8.9\e{46} \ \rm s^{-1}$  H I atoms crossing the shock per second at an age of 18 years,  corresponding to a density of $n({\rm H I}) \sim 60-70 \  \ccm$.  From the X-ray luminosity they estimate an ionizing luminosity of $3.7\e{45} \ \rm s^{-1}$ by number, corresponding to $\sim 8\e{34} \ \ergs $. They therefore estimate that $\sim 4 \%$ of the incoming hydrogen atoms become ionized as they cross the shock. 

The observations by \cite{Smith2005}  only extended to
 2005 Feb. (day 6577), and had rather
large observational uncertainties arising from the different instruments and lines used,
limiting the extent to which firm conclusions with regard to this
prediction could be made. \cite{Heng2006} used available STIS observations and
showed that H$\alpha$ had increased by a factor of $\sim 4$ from 1997
to 2004. Unfortunately, there was little spatial overlap between the
various epochs and the uncertainties in the flux were therefore large.

In spite of not covering the
full ejecta because of the narrow slit, our UVES observations in Fig. \ref{fig4}  have the advantages of
being obtained with the same instrumental setup, as well as monitoring
the same area of the ejecta and ring. In addition, our time coverage is a factor
of two longer. Further, the FORS2 observations in Fig. \ref{fig4a} cover the full ring (except for minor slit losses due to the seeing). These provide a good estimate of the total flux, as well as a check of the evolution as determined from UVES. 

The light
curve of the reverse shock (Fig. \ref{fig_core_rev}) exhibits a steady increase in the
flux by a factor of $\sim 3.5$ between days 5000 and day 7500. 
 However, after $\sim $7500
days there is a clear flattening of the light curve. The total increase from $5000 - 9000$ days is a factor $4-5$.
We note that also the monochromatic flux of $\Ha$ at $-$3000 \kms \ in Fig. \ref{fig5b} shows a similar flattening. This velocity is dominated by the reverse shock and is subject to less systematic errors from the interpolation between the lines.  

A similar flattening as seen in the reverse shock flux evolution has been seen in other aspects of the ring collision. In the optical range the flux evolution of the narrow lines from the shocked ring show a flattening of most emission lines from day $\sim 7000$ \citep{Groningsson2008b}. These lines are formed behind the slow shocks transmitted into the ring, mainly as a result of the photoionization by the soft X-rays from these shocks. Their flux is therefore a measure of the ejecta -- ring interaction. 

In the X-ray range, 
\cite{Park2011} find a similar flattening in the 0.5 -- 2 keV X-ray flux.The hard 3 -- 10 keV flux, however, continues to increase. Decomposing the soft flux within the reflected shock model by \cite{Zhekov2010}, Park et al. find a find a dramatic decrease of the
X-ray flux produced by the shock that is transmitted into the ring. It is not clear that this is consistent with the evolution of the optical lines.  The degree of this flattening has, however, been challenged by \cite{Maggi2012}. The radio emission, correlated with the hard X-rays, continues to increase up to at least 2010 (day 8014) \citep{Zanardo2010}.

The ionization of the hydrogen close to the reverse shock depends on both the ionizing flux and the density. The latter can be estimated from models of the ejecta used for light curve calculations and spectra during the first years after explosion. 
From the  \cite{Shigeyama1990} 14E1 model we can approximate the density profile at 20 years with
 $\rho(V) = 2.0\e{-23} (t/20 \ \rm yrs)^{-3} \ V_4^{-8.6} \ \gccm$. \cite{Michael2003} fit the 10H model of \cite{Woosley1988} with a similar power law, but the density is a factor of 3.0 higher. The uncertainty in the density is considerable, both from differences in the explosion models and also from the variations in the exact position of the reverse shock. We therefore introduce a factor $k$ to take this into account, which is expected to be in the range $1 \la k \la 3$. 
With an H : He abundance of 1 : 0.25 by number we get for the envelope density
\begin{equation}
n_{\rm H \ I}(r)= 71 k  \left({r \over R_{\rm s}}\right)^{-8.6} \left({t  \over 20 \ {\rm yrs}}\right)^{5.6} \ \ccm.
\label{eq_hiden}
\end{equation}
where $R_{\rm s}$ is the radius of the reverse shock. The hydrogen density at the reverse shock therefore increases rapidly with time as long as the ejecta are in the steep part of the density profile (i.e. $V \ga 4000$ \kms). At lower velocity the density gradient flattens considerably, and consequently the density at the reverse shock will increase more slowly and ultimately decrease as the flat part of the density profile is encountered for $V \la 3000 - 4000$ \kms. This will, however, only occur at $\ga 40$ years. This assumes an ejecta structure similar to that of the 1D models discussed above, and instabilities may greatly change this (see below).  

From  2005 Jan. (day 6533), which was the epoch of the observations by \cite{Smith2005}, to  2010 Sept.  (day 8619) the 0.5 - 2 keV X-ray flux has increased by a factor of $\sim 3.6$ \citep{Park2011}. 
From Eq. (\ref{eq_hiden}) we, however, find that the density at the reverse shock during this period has increased by a factor of $\sim 4.7$. The ionization of the pre-shock gas should therefore be nearly constant, or even decreasing in the last epochs. This applies especially to the period when the X-ray  flux levels off. We do therefore not expect a rapid decrease of the flux from the reverse shock as was predicted on the basis of a steadily increasing X-ray flux.

In the case of a stationary reverse shock, as may be the case for the shock in the ring plane, we expect the intensity of the H$\alpha$ line to be  
\begin{equation}
I_{\nu}  \approx  {\epsilon \over 4 \pi} n_{\rm H I}(R_s)  V_{\rm ejecta}(\rm R_s)
  \label{eq_inu}
\end{equation}
where $\epsilon = 0.2 - 0.25$ is the number of $\Ha$  photons per ionization
between the shocked electrons and the free-streaming H I and $V_{\rm ejecta}(\rm R_s) $ is the ejecta velocity at the shock \citep[e.g.,][]{Michael2003}. With the density from  Eq. (\ref{eq_hiden}) we find that $I_{\nu}  \propto   t ^{5.6} V_{\rm ejecta} \propto   t ^{4.6}$  \citep[see e.g.,][]{Heng2006}. From Fig. \ref{fig_core_rev} we find that the flux of the reverse shock has increased by a factor $\sim 4.5$ from day 5040 to day 7170. After this it stays nearly constant. From the above scaling we would expect it to have increased by a factor of  $\sim 5.1$, i.e., close to what is observed.  

The scaling  above only applies to a stationary shock of constant surface area. In reality, the reverse shock is expected to expand (in the observer frame) above the ring, where it sweeps up the constant, low density H II region. In this case \cite{Heng2006} find that the flux should increase as $F \propto t^{(2n-9+4s-ns)/(n-s)}$, where $n$ is the power law index of the ejecta density and $s$ that of the circumstellar medium. For $n=8.6$ and a constant density medium, we find $F \propto t^{0.95}$. The observed flux should therefore evolve somewhere between these limits, although the ring plane should dominate and a flux evolution closer to $F \propto t^{4.6}$ is expected, which also gives the best agreement with the observations up to $\sim 7100$ days. 

The nearly constant flux after $\sim$ 7000 days may be a result of an early encounter with the higher density inner hydrogen region, close to the core. The flux evolution will then be determined by the area involved in this and the density, $F \propto \Omega(t) \  \rho_{0}(V_{\rm ejecta}(R_s)) \  t^{-3} $, where $\Omega(t)$ is the solid angle of the 'high density' encounter and $\rho_{0}$ is the density at a reference  time $t_0$. 
From the HST observations in L13 it is indeed found that there is especially in the southern part of the ring a bright region in H$\alpha$, which may come from such a high density region of the ejecta. At the last HST observation discussed in L13 at 8328 days (22.5 yrs) this feature is close to the projected position of the reverse shock. Based on its low radial velocity it should be close to the plane of the sky, rather than the ring plane. Taking a maximum radial velocity of 1500 \kms, the position corresponds to an ejecta velocity of $\sim 4500 - 5000$ \kms. It is therefore likely that the ejecta has a density at lower velocities considerably different from the 1D models. This is also confirmed by the simulations by \cite{Hammer2010}. An evolution of the reverse shock flux departing from the above simple scalings should therefore not be surprising.

The line profiles combined with the spatial information provide important constraints on the reverse shock geometry, as was demonstrated by  \cite{Michael2003}. This applies especially to the maximum velocity, which is limited by the equatorial ring in the ring plane.  

Our late spectra can give additional information on this from the high velocity wings of  H$\alpha$ . 
 \cite{Sugerman2005} find a semi-major axis of the ring of 0.82\arcsec \ and an inclination of 43\degr. This gives a radius of the ring equal to $6.1\times10^{17}$ cm for a distance of 50 kpc. Assuming a  radius for the reverse shock equal to $\sim 80$\% of the blast wave (ring) \citep{Chevalier1982} the radius of the reverse shock is $\sim  4.9\times10^{17}$ cm, and the ejecta velocity at the shock therefore
 \begin{equation}
V_{\rm ejecta}(R_{\rm s}) = 7740 \left({t  \over 20 \ {\rm yrs}}\right)^{-1} \ \rm km \ s^{-1} \ ,
\end{equation}
where $t$ is the time since
explosion in years. The maximum LOS
velocity of the ejecta in the ring plane is therefore expected to be
\begin{equation}
V_{\rm max} =
5,660~  \left({t  \over 20 \ {\rm yrs}}\right)^{-1}   \kms. 
\label{eq_vmax}
\end{equation}

The fact that we observe velocities up to $\sim
11,000$ \kms \ for H$\alpha$ even in our last 2009 observations
(Fig. \ref{fig4}), 
suggests that %% the ejecta is still expanding fairly unaffected by the
%% CS environment outside of the ring plane and that
the expansion of the
ejecta must now be anisotropic, with a larger extension
above the ring and near stagnation in
the plane of the ring.

It is interesting to compare our observed $\Ha$ expansion velocity
with the morphology found from the modeling of radio observations, as well
as $\Lya$ observations. \cite{Ng2008} find from the radio emission
that the best fit to the morphology is provided by a torus model with
a thickness of $\pm 40 \degr$from the equatorial plane, which is
similar to the thickness found by \cite{Michael1998a,Michael2003} from
the $\Lya$ modeling, $\sim \pm 30\degr$. 
With an inclination of $\sim
43 \degr$ of the ring, this means that the reverse shock extends up to nearly
the direct LOS to the center of the supernova. A maximum observed velocity of
$11,000$ \kms \ therefore also corresponds to a similar radial
velocity in this direction, and an extension of the ejecta to $\sim 6.9
\e{17} (t/20 \ \rm yrs)$ cm, or $\sim 15 \%$ larger than the radius of
the ring and $\sim 40$ \% larger than the reverse shock location in the ring plane.
That the maximum velocity comes from the part of the reverse shock expanding in the LOS is confirmed by the STIS observations in Fig. \ref{fig6bb}, which shows the highest velocity to occur close to the center of the ejecta.

We find little change in either the maximum velocity or the flux of
the red and blue wings above $\sim 8000$ \kms \ (Figs. \ref{fig4} and
\ref{fig4a}). The maximum velocity is, however, difficult to determine accurately because of  blending with the [O I] lines from the ejecta on the blue side and narrow lines on the red. 
The STIS observations in Fig. \ref{fig6bb}, however, show that there does seem to be a decrease of the reverse shock velocity in H$\alpha$ close to the LOS of the center of the ring. Although contaminated by the [O I] emission from the ejecta especially in 2004, we find in the  2004 spectrum a maximum blue velocity of $ -12,200$ \kms, while in 2010 this velocity has decreased to  $-9100$ \kms. The corresponding values on the red side are $+8300$ \kms, and $+7400$ \kms. The velocities on the red side are  lower than the maximum measured with UVES and FORS2, which are likely to be a result of the limited S/N of the STIS spectra. 

The steepening of the line profile with time is
roughly consistent with the maximum LOS velocity of the ejecta
in the plane of the equatorial ring (see Sect. \ref{sec_rev_shock}). This is the point
of maximum interaction, and it is therefore not surprising to find that the
most dramatic change  occurs around this velocity, as was concluded already by
\cite{Michael2003}. If the interaction takes place in a thin
cylinder concentric to the ring with small extent in the polar
direction, and with roughly equal strength along the rim, it will
result in a double peaked line profile between $\pm V_{\rm max}$ 
(Eq. \ref{eq_vmax}). As the thickness of the shell increases in the
polar direction the line approaches a flat, boxy profile
\citep[see][for a discussion]{Michael2003}. The increasing flux of the
central peak of the line is therefore partly a result of the increase
of this component. As Figs.  \ref{fig_core_rev} and \ref{fig5b} show, there is, however, also a substantial increase of the flux from the inner ejecta component on top of this.

The fact that there is emission above $\sim 8000$
\kms \ may require the ejecta to interact with a medium of
comparable density to that in the plane. 
 At 20 years and
     11,000 \kms \ we get from Eq. (\ref {eq_hiden}) a number density $n \approx 3$
     cm$^{-3}$. With this density profile we find a ratio between the
     ejecta density in the ring plane, where $V_{\rm ej} \approx
     7740$ \kms, and the density at 11,000 \kms \ equal to $\sim 20$. With
     an H II density of 100 cm$^{-3}$ \citep{Chevalier1995} an estimate based on
     $V_{\rm s} \approx (\rho_{\rm ej}/\rho_{\rm H II})^{1/2} V_{\rm
       ej}$  gives a shock velocity of $\sim 1900$ \kms \ into the H
     II region above the ring plane. 
     
According to \cite{Mattila2010}, 100 cm$^{-3}$ H II region gas can, however, not fill the full volume from the ring plane up to our LOS, as this would violate observational constraints on the narrow line emission. It is therefore likely that the H II region is less dense along our LOS than 100 cm$^{-3}$, which results in a higher shock velocity than 1900 km/s. This is also indicated by 
\cite{Ng2008}  who find from the radio a mean expansion speed of $4000 \pm
400$ \kms \ over the period 1992 -- 2008. However, X-ray observations
by Chandra indicate a considerably lower expansion speed of $\sim
1625$ \kms  \  \citep{Park2011}.  

The high velocity emission from the ejecta in the LOS could also be a result of the X-ray ionization from the ring. If a large fraction of the X-ray flux is below $\sim 0.1$ keV these X-rays may be absorbed in the outer ejecta and at high altitudes from the ring (see Fig. \ref{fig_endep_2d} below). They can there provide the necessary energy for the H$\alpha$ emission. The fact that there is little change in the flux in the high velocity wing of H$\alpha$, however, argues against this explanation. 

Adopting a similarity solution we expect $V_{\rm ej} \propto
t^{-1/3}$, giving a decrease in the ejecta velocity by
$\sim 20 \%$ over 10 years. There is reason to doubt that the similarity
solution applies over the small time range since the impact on the H
II region.
With a maximum velocity of 11,000 \kms \ this may be in conflict with the
observations (Fig. \ref{fig4}). As pointed out in Sect. \ref{sec_hydrogen}, the maximum velocity could be higher and a decrease would then be 'hidden' in the many lines from the ring collision.

\subsection{Ejecta component}

Based on photometry of the ejecta with the WFPC2, ACS and WFPC3 on HST,
L11 find that the flux has increased by a factor of 2
-- 3 in the R- and B-like bands from day 5000 to day 8000. The R-band 
is dominated by $\Ha$, while the B-band is a mix of $\Hb$
and Fe I-II lines \citep{Jerkstrand2011}. 

Our light curve of  $\Ha$
(Figs. \ref{fig_core_rev} and \ref{fig5b}) shows an increase by a factor of $4-6$, while the [Ca II] \wll 7292, 7324 lines
(Fig. \ref{fig7}) increased by a factor of $5 - 8$ over the same period. Given that the HST observations cover a wide spectral range for each filter and that there is a considerable uncertainty in the line fluxes because of blending with the narrow lines, this is consistent with the increase found from the HST broad band photometry.  This therefore confirms spectroscopically what was concluded based mainly on imaging in L11.  In addition, we see that other lines, in particular the Mg II \wll 9218, 9244  lines (see Fig. \ref{fig_ejecta_spectra}), show a similar, or even larger increase. 

Later than $\sim $ 7000 days  the $\Ha$,  [Ca II]  and Mg II fluxes level off, similarly to the reverse shock lines and the X-rays, discussed in the previous section. As we discuss in next section (see also L11),  there is a close connection between the X-rays from the ring collision and the ejecta flux. It is therefore expected that the ejecta flux follows the  flattening seen in X-rays. 

As for the line profiles from the ejecta, $\Ha$ is severely contaminated by the narrow lines, as well as the reverse shock emission. The [Ca II] lines are less affected by these and also lack a reverse shock component (Fig. \ref{fig6b}). The line shape on the blue side is well defined and approximately  triangular, similar to that of the $\Ha$ line in the early STIS observation in Fig. \ref{fig_ha_1999_2004}. The blue side reaches $\sim 4500$ \kms, and does not change appreciably from 2002 to 2012 (days 5738 -- 9019). This velocity is similar to $\Ha$, as measured from the 1999 STIS G750M spectrum in Fig. \ref{fig_ha_1999_2004}. Both the similar line profiles and the similar maximum velocities indicate that these lines arise from the same region. 
 
L13 show that the
changing morphology results from the same cause as the increase observed in
the light curve. They find that the $\Ha$ emission from the inner ejecta has changed from a centrally dominated, elliptical profile before year $2000$ to one dominated by a ring-like structure at later stages.  Given this complex morphology it is somewhat surprising that the line profiles from the core component change relatively little, except in flux (see Figs. \ref{fig4} and \ref{fig_ha_1999_2004}). 
As noted above, a reason for this may be that most of the change occurs for LOS velocities less than $\sim 1500$ \kms, a spectral region which as discussed earlier is blocked by the narrow lines.  

There is also little evolution of the [Si I]/[Fe II] line from the VLT/SINFONI observations (L13). This emission is, however, centrally peaked and is likely to come from more central, processed material, protected from most of the X-rays, and does not display any major change in morphology between 2005 and 2010. 

\subsection{The energy deposition by the X-rays from the ring collision}
\label{sect_xrays}
L11 show that up to day 5000 the light curves in the R
and B-bands are compatible with the radioactive decay of \iso{44}Ti.
They concluded that to explain the increasing ejecta flux an additional
energy source is needed. The most likely such source is the soft
X-rays emitted by shocks resulting from the ejecta - ring
interaction.  We now discuss the details of the deposition of these and their conversion into optical/UV radiation. 

The observed X-ray luminosity above 0.5 keV is $\sim 4 \times 10^{36} \ergs$ at the last observation at  8619 days by  \cite{Park2011}. Based on the spectra, \cite{Zhekov2010} has modeled the X-ray emission  with three
components, one fast blast wave, one soft component from the transmitted
shocks into the dense clumps in the ring and one from the reflected shock
component. The soft radiation from the transmitted shocks into the ring, with $k T_e \sim
0.35$ keV,  dominates the
total flux.  

After $\sim 10$ years the hydrogen envelope was mostly
transparent to X-rays with energy $\ga 1$ keV (L11). The inner core region,
containing both mixed-in hydrogen and synthesized heavy elements, which increase the opacity, is, however,  still opaque up to $\sim 8$ keV. The X-rays will there be
thermalized into optical, IR and UV radiation. Only after $\ga 30$
years does the whole core become transparent.

The exact X-ray luminosity and spectrum is uncertain. 
Observations only probe energies down to $\sim 0.5$ keV. 
For low shock velocities most of the
X-ray flux is below the low energy interstellar X-ray cut-off at $\sim 0.5$
keV. In \cite{Groningsson2008a} it is shown that the 
shocked lines have a  FWHM $\sim 200$ \kms. Even if
there could be an inclination effect, which could increase this velocity by a factor
up to $\sim 1.4$, the typical shock velocities giving rise to these lines
are $\la 300$ \kms, corresponding to shock temperatures $\la 0.1$
keV. In addition, \cite{Groningsson2008b} show that shocks into the densest parts of the ring are radiative up to $\sim 500$ \kms, or $\sim 0.25$ keV. Therefore, a large fraction of the X-rays/EUV flux may well be below
the observed ISM absorption threshold at $\sim 0.5$ keV.

The X-rays give rise to secondary
electrons by photoionization in the same way as the gamma-rays and
positrons from radioactivity \citep{Shull1985, Xu1991,  Kozma1992}. These in turn will lose their energy in non-thermal excitations,
ionizations and Coulomb heating of the free thermal electrons. The
resulting optical/UV spectrum from the X-ray input will therefore not be very different from
that of the radioactive decay, as long as the ionization of
the gas is low. For a low ionization each hydrogen
ionization requires $\sim 30$ eV at the relevant level of
ionization \citep{Xu1991}. 
Once the ionization increases above $\sim 10^{-2}$
the efficiency of excitation and ionization decreases, while that of
the heating increases. The cooling, balancing the heating, is close to the core region at these epochs mainly done by thermal, collisional excitations of
mid- and far-IR fine structure lines and molecules, and possibly cool
dust. This emission will therefore be difficult to observe directly, although far-IR and mm observations may help. In the envelope adiabatic cooling dominates.

Compared to the radioactive excitation some important differences arise due to the fact that the X-ray illumination is from the outside, while
that of the gamma-rays and positrons is from the central regions.
The positrons from the
\iso{44}Ti decay, which dominate the radioactive energy input at these epochs, will
 be deposited locally, mainly in the Fe-rich gas, unless the
magnetic field in the ejecta is extremely low \citep[see discussion in][]{Jerkstrand2011}. The X-rays will on the
other hand penetrate  to different depths,
depending on their energy.

The photoelectric energy deposition is sensitive to the density and abundance distributions in the ejecta. To calculate this we use the ejecta model, 14E1 from \cite{Shigeyama1990}, mixed as in  \cite{Blinnikov2000}. This is a spherically symmetric model, where the chemical mixing has been modified to reproduce the light curve. In reality, Rayleigh-Taylor instabilities will mix the different nuclear burning zones in velocity, resulting in both large abundance and density inhomogenites \citep[e.g.,][]{Hammer2010}. Nevertheless, this model should show the main qualitative features of the X-ray deposition.

 Using the same code as was used in L11 for calculating the total energy deposition of the X-rays, we can calculate also the distribution of the energy deposition in radius.  We assume here that the X-ray sources are located in the plane of the equatorial ring at $6.1\e{17}$ cm and the ejecta extend to the reverse shock at $  R_{\rm s} = 4.9\times10^{17}$ cm. In addition, we assume spherical geometry for the ejecta, which is certainly an oversimplification. 

\begin{figure}[!h]
\resizebox{\hsize}{!}{\includegraphics{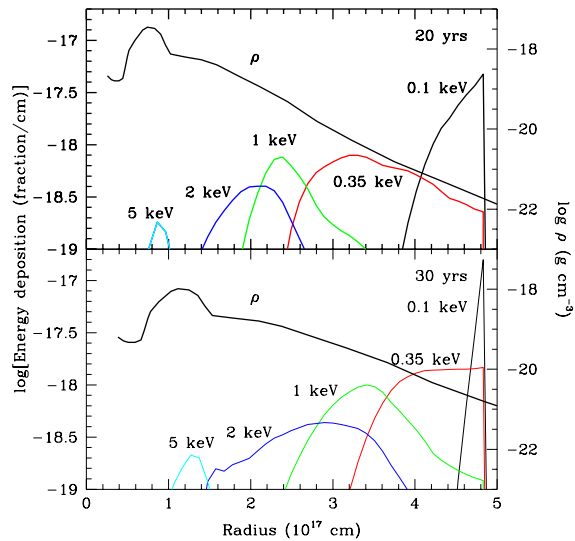}}
\caption{Energy deposition per cm for the mixed 14E1 model for different photon energies at 20 years and 30 years after explosion.  We also show the density distribution from the 14E1 model 
from \cite{Blinnikov2000} scaled to these epochs. Note the sharp density gradient in the core at 
$r  \sim   1.0\e{17}$ cm and the power law gradient for $r  >  1.5\e{17}$ cm (both at 20 years), corresponding to $\sim 4000$ \kms. The velocity at the ring is 10,500 \kms \ at 20 years and 6900 \kms \ at 30 years.}
\label{fig_endep}
\end{figure}

In Fig. \ref{fig_endep} we show the fractional energy deposition per cm integrated over the polar angle, $df/dr$, for energies between 0.1 -- 5 keV at 20 and 30 years after explosion. The total energy deposition in a shell with radius $r$ and thickness $dr$ is then $dL(E) = L_{\rm X-ray}(E) \ df(E)/dr \  dr$.  As we show below, this is, however, only a rough approximation and the real deposition function is two dimensional or even three dimensional. 

We here see that for the hard X-rays with $\ga 1$ keV most of the energy is deposited at the outer boundary of the core, while the softer X-rays are deposited in the hydrogen envelope. At 30 years the soft X-rays are deposited in a more narrow velocity interval than at 20 years, caused by the increasing ejecta density close to the reverse shock (Eq. \ref{eq_hiden}). 
The harder X-rays, however, are deposited deeper in the core, where the flatter density distribution does not have the same effect.  The fact that the core is very heterogeneous in abundance, with a large fraction of the volume occupied by the iron bubble \citep{Li1993}, having a large X-ray opacity, means that most of the hard X-rays will be absorbed efficiently here.  

\begin{figure}[!h]
\resizebox{\hsize}{!}{\includegraphics{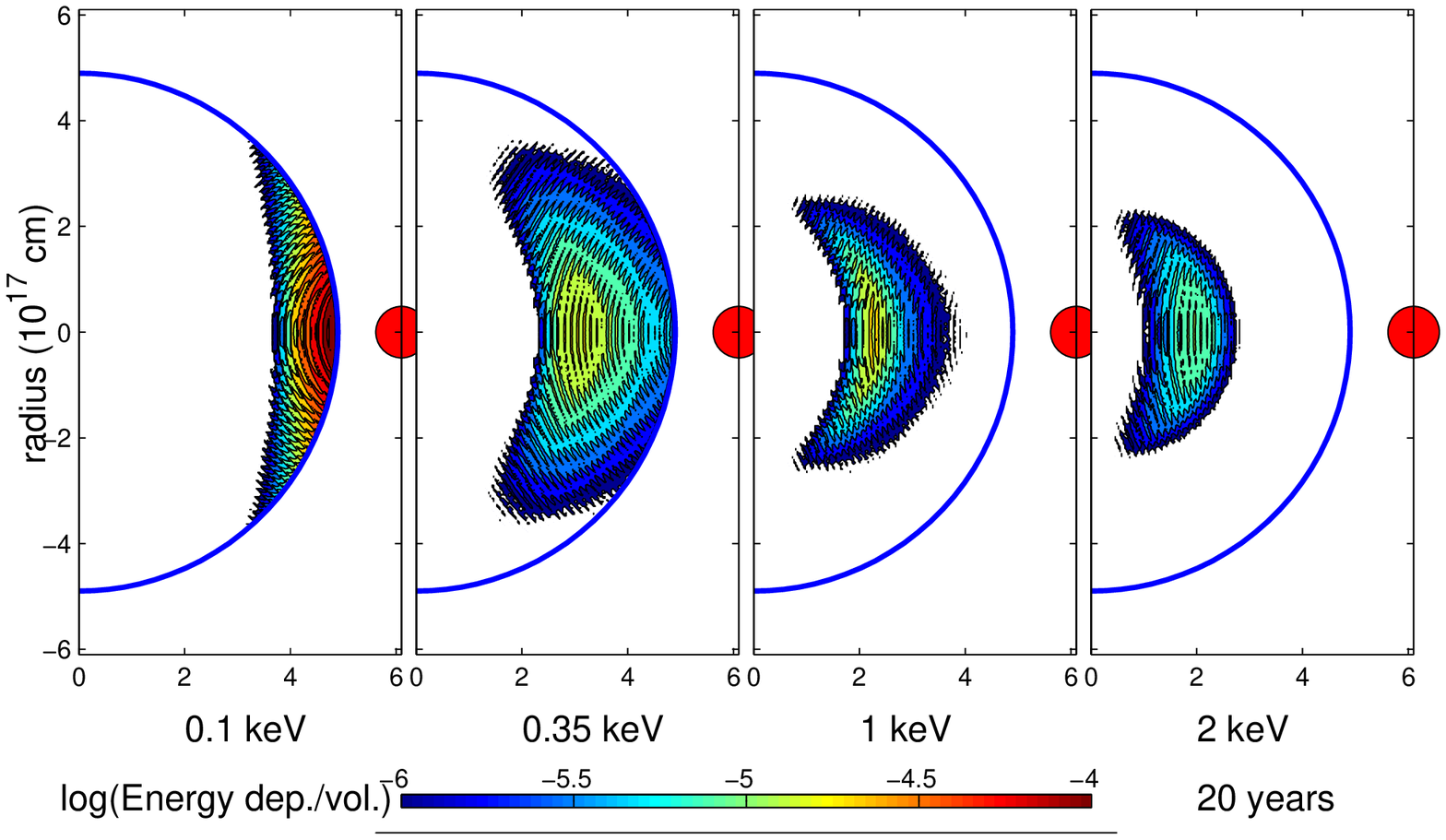}}
\resizebox{\hsize}{!}{\includegraphics{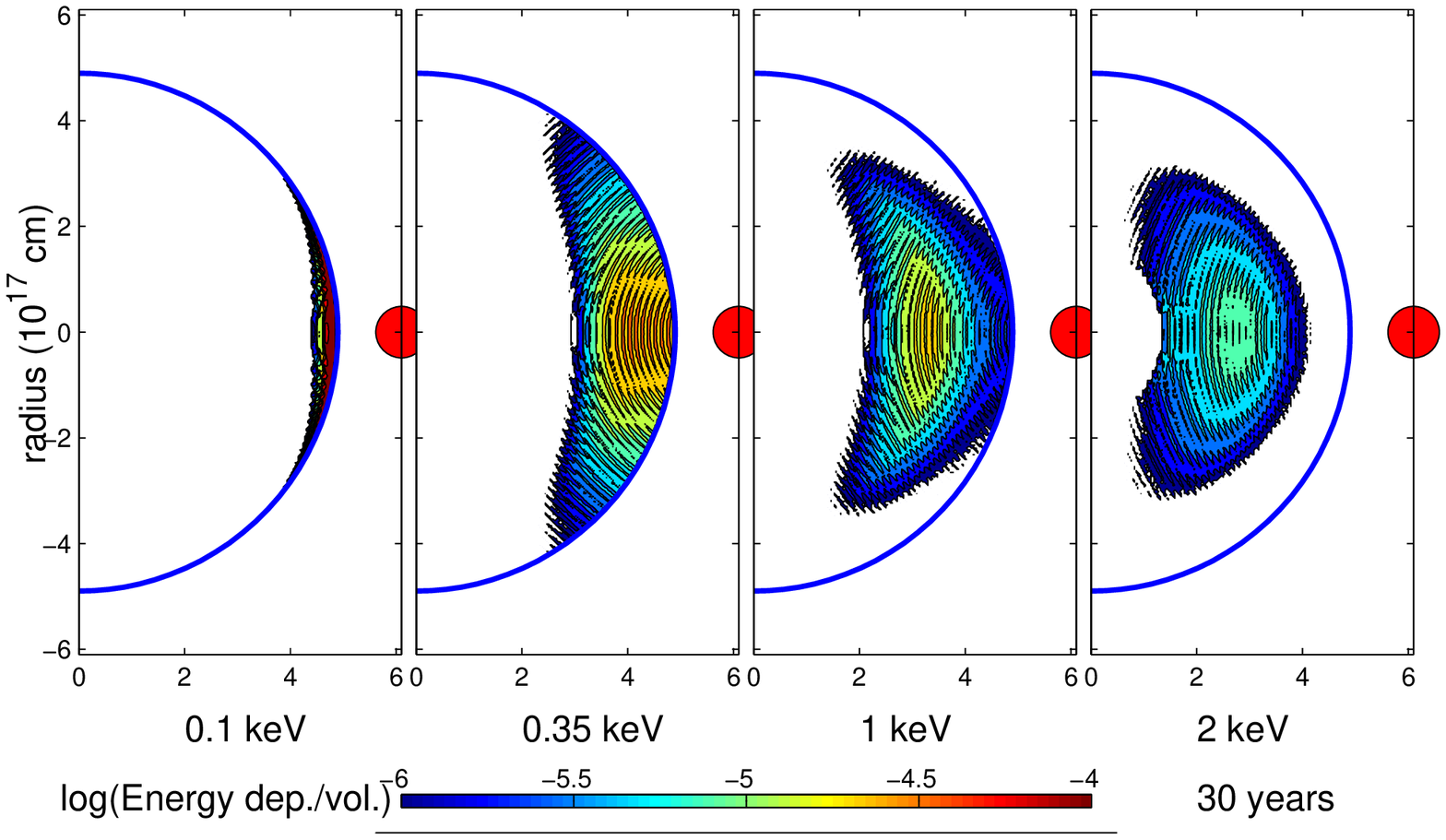}}
\caption{Energy deposition for the same model as in Fig. \ref{fig_endep} but now shown perpendicular to the equatorial plane at 20 years (upper panels) and 30 years (lower panels) after explosion. The x- and y-scales are in $10^{17}$ cm. The red circle at $6.1 \times 10^{17}$ cm marks the position of the source of X-rays (i.e., the ring), while the blue circle with radius $4.9 \times 10^{17}$ cm marks the location of the reverse shock, assuming spherical symmetry. The different contours and colors are distributed according to the logarithm of the energy deposition per volume, shown in the color bar at the bottom of each panel.  }
\label{fig_endep_2d}
\end{figure}

Because the interaction of the ejecta and ring is mainly taking place in a thin region around the equatorial ring, the X-ray illumination of the ejecta will be highly non-uniform in the polar direction. Most of the energy deposition will occur close to the equatorial plane, but especially at higher energies a large fraction of the ejecta will be transparent and the deposition more distributed also above the plane. To calculate the deposition in 3-D we assume that the X-rays are injected in a thin region in  the equatorial plane, and azimuthally uniform. This is a reasonable approximation for observations obtained after $\sim 2004 $, but for earlier epochs the distribution of the individual hotspots in the azimuthal direction should be taken into account. In Fig.  \ref{fig_endep_2d} we show the deposition at different energies in the poloidal plane.

Again, we note the larger penetration of the hard X-rays. More interesting is, however, that the X-rays are deposited spatially fairly locally. At low energies this is close to the ring and also close to the equatorial plane, while the higher energies are deposited in a more extended region closer to the core.  We also note the higher concentration of the energy deposition to the ring at 30 years, compared to 20 years.

As mentioned in L11, the restricted radial range of the energy deposition may affect the morphology of the HST images, which show a ringlike structure in the ejecta. This may be explained as a result of the limited penetration of the X-rays from the shock, and therefore reflects the energy deposition in the ejecta rather than the true density distribution, as was seen in Fig. \ref{fig_endep}. This is discussed in more detail in L13. 

Figure \ref{fig_endep} also shows that if most of the X-ray luminosity is below $\sim 0.5$ keV, as may well be the case, the outer parts of the ejecta may absorb a larger fraction of the energy, while the core only receives a minor fraction of the X-ray input. A further brightening of the outer parts can therefore be expected. As has been mentioned earlier, recent observations by \cite{Park2011}, however, show a flattening of the soft (0.5 -- 2 keV) X-ray flux (see however \cite{Maggi2012}), while the hard (3 -- 10 keV) flux continues to increase. 

In the high velocity hydrogen envelope the density is 
low, and the recombination timescale long.  For a density
profile similar to that in Eq. (\ref{eq_hiden})  the density at the reverse shock is $70 - 200 \ \ccm$ at an age of 20 years. In the absence of X-ray input we find using the code in \cite{2002NewFransson} that the temperature is here expected to be $\sim
10-20$ K . Extrapolating the Case B recombination rate from \cite{Martin1988} gives a recombination rate $\sim 2
\e{-11} \  \rm cm^{3} s^{-1}$ at 20 K and a recombination time
\begin{equation}
t_{\rm rec} = 22 \ k^{-1} x_e^{-1}  \left({r \over R_s}\right)^{8.6} \left({t  \over 20 \ {\rm yrs}}\right)^{-5.6} \ \rm years, 
\label{eq_rect}
\end{equation}
where $x_e$ is the ionization fraction. Equation (\ref{eq_rect}) shows that the recombination time decreases fast both with time and with decreasing radius. As long as $x_e \ll 1$ it will, however, be long at least close to the shock. As the ionization increases  recombination becomes more and more important.

From the same calculation as above, again including only the radioactive input, we find that
the degree of ionization in the envelope is nearly constant with time
at $x_e \approx 7\e{-3}$, as the ionization is frozen-in. Because of the X-ray illumination the ionization may, however, increase considerably. Because of this and the large density gradient, the recombination timescale may become short only a short distance inside the reverse shock. A stationary situation may then be at hand where absorption and emission balance. 

For the regions where the recombination timescale is long, the ionization of the
envelope is  determined by the number density of H atoms in the part of the 
envelope where the X-rays are deposited and the time
integrated ionizing luminosity, rather than the instantaneous
luminosity. Further in, once the ionization has increased enough there may be a balance between ionizations and recombination.

Integrating the X-ray luminosity from \cite{Park2011} over time gives a total
X-ray energy $E_{\rm X-ray} \approx 1.6 \times 10^{44}$ ergs in the
0.5-2 keV band. Two thirds of this is emitted in the period 19 -- 22
years after explosion. On top of this there is a minor contribution
from hard X-rays and a possibly dominant contribution below 0.5
keV. Roughly half of the X-ray energy will be emitted inwards to the ejecta. As shown above, most of the X-rays and EUV photons below
$\sim 0.3$ keV are absorbed by the envelope. The luminosity below this
energy is therefore most important for the ionization of the envelope,
but also the most uncertain observationally. 

To take into account the fact that we do not know the level of the unobserved soft X-ray flux, we assume that the total time integrated luminosity emitted in a narrow energy interval in the EUV  is
 $E_{\rm EUV} = f  \times 10^{44}$ ergs, where $f$ is a scaling factor reflecting the uncertainty in the EUV flux. 
To get an approximate estimate of the ionization in the ejecta we further assume that recombinations are slow and that each volume is absorbing a fraction of $E_{\rm EUV}$ given in Fig. \ref{fig_endep_2d}. We also ignore the decreasing efficiency of ionization as the ionization increases. Finally, for simplicity we ignore the expansion, which is justified by the fact that most of the X-rays are emitted during a  period short compared to the age of the supernova. These assumptions affect the quantitative results, but hardly the qualitative. With these assumptions the ionization of a given volume depends on the deposited energy and is inversely proportional to the density. 

\begin{figure}[!h]
\resizebox{\hsize}{!}{\includegraphics{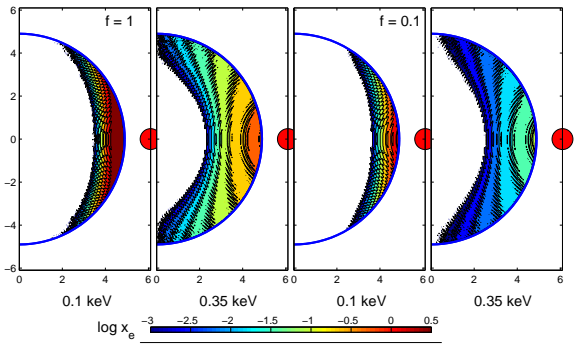}}
\caption{Ionization for the same model as in Fig. \ref{fig_endep_2d} at 20 years after explosion. The different contours and colors are distributed according to the logarithm of the fractional ionization, $x_e$, shown in the color bar at the bottom of each panel.  The two sets give the ionization for two different assumptions of the ionizing flux.  The x- and y-scales are in $10^{17}$ cm.}
\label{fig_ion_2d}
\end{figure}

In Fig. \ref{fig_ion_2d} we show the resulting distribution for two values of the efficiency of ionization, $f=1$ and $f=0.1$, which may bracket the range. Not surprising, the distribution reflects the energy input in Fig. \ref{fig_endep_2d}. More interesting is that close to the shock the ionization is expected to be fairly high, in particular for the case where most of the ionizing flux is below $\sim 0.3$ keV. Comparing the 0.1 keV and 0.35 keV cases, we note the higher, but more concentrated ionization in the former case for the same total energy.

If the recombination time close to the reverse shock is long most of the emission in the H I lines comes from direct excitation of these. The fraction of the deposited energy going into excitation of levels with $n > 2$ is $\sim 12 \%$ \    for $x_e = 10^{-2}$  \citep{Xu1991}. Of this $\sim 10 \%$ \  is emitted as $\Ha$ emission. Assuming that we underestimate the UVES luminosity by a factor of $\sim 2$, as was found from the 2004 STIS observation (Sect. \ref{sec_stis}), our observation at $\sim$ 8000 days (Fig. \ref{fig_core_rev}) result in  $L({\rm H}\alpha) \approx 3.2\e{34} \ \ergs$.
To explain this with direct excitation only, an X-ray luminosity of $\sim 10^2 L({\rm H}\alpha) \approx 3.2\e{36} \  \ergs$ is therefore needed. In the regions where the recombination time may become short compared to the age also the non-thermal ionizations and subsequent recombinations will contribute to the $\Ha$ luminosity, increasing this by a factor of $\sim 3$, compared to the direct excitations. The observed 0.5 -- 2 keV X-ray luminosity at 8000 days was $\sim 3.6 \times10^{ 36} \ergs$\  \citep{Park2011}. An additional $\sim 14 \%$ was in the hard 3 - 10 keV range. There does therefore seem to be enough X-ray luminosity to explain the observed $\Ha$ luminosity.

The conclusion of this discussion is  that the ionization in
both the envelope as a whole and the core are likely to remain low for
at least the next ten years. A dramatic change of the spectrum is
therefore not expected, although the line ratios may slowly change. 

The estimate by \cite{Smith2005} of the ionization assumes that most of the X-rays are absorbed close to the reverse shock. The H I density in front of the reverse shock is at 20 years $\sim 90 \ \ccm$. The mean free path close to the Lyman limit is then $\sim  1.6\e{15} (E/13.6 \ {\rm eV})^3 (n/100 \ \ccm)^{-1}$ cm, so most of the ionizing photons are indeed absorbed close to the reverese shock as long as the photon energies are not too high. As Figs.  \ref{fig_endep}  and  \ref{fig_endep_2d}  show, this is in general the case for photon  energies $\la 0.1$ keV, but not for higher energies. 
The fact that we still observe both H$\alpha$ and Ly$\alpha$  shows that the X-rays are not yet able to pre-ionize the gas completely. As we see from Fig. \ref{fig_ion_2d}, the ionization decreases substantially above the ring plane, and at least a fraction of the H$\alpha$  may originate here.

Except for the Mg II \wll 9218, 9244 lines, there are  no major
changes in the relative fluxes of the lines from the core
(Fig. \ref{fig_ejecta_spectra}). The Mg II lines, however, increase substantially compared to e.g., the [Ca II] lines.   If these lines are pumped by Ly$\alpha$ there are several factors which influence the outgoing flux. The most straightforward is  the increasing Ly$\alpha$ flux, which should be reflected directly in the Mg II lines. However, also the fractional population of Mg II, which determines the optical depth of the Mg II lines, and therefore the pumping efficiency, may increase because of the ionizing effects of the X-rays. Finally, the fraction of the envelope which is in velocity resonance with the pumping transition may change with time.

\subsection{Dust}

Strong evidence for early dust formation in the ejecta was found from both blueshifts in the line profiles \citep{Lucy1989} and a thermal IR excess \citep{Wooden1993}. The last far-IR observations before the ring interaction are from
day 1731, where \cite{Bouchet1996} find a dust temperature of $\sim$155 K. There is also from the line profiles of  Mg~I] \wl4571 some evidence for dust  from the HST observations at 8 years with an effective optical depth close to unity \citep{Jerkstrand2011}. This is, however, likely to be in the form of very dense clumps with high optical depth. The effective optical depth is therefore only a measure of the covering factor of the dust clumps.

More recently \cite{Matsuura2011} have from Herschel far-IR observations found evidence for a large dust mass, estimated to $0.4 - 0.7$ \msun, but could be up to $2.4$  \msun. The mass estimate is, however, highly dependent on the dust temperature and the mass could be several orders of magnitude lower if the temperature is higher than adopted by \cite{Matsuura2011}, as discussed from APEX observations by  \cite{Lakievic2012}.

As was discussed in Sect. \ref{sec_stis}, the 1999 $\Ha$ line profile may indicate a $\sim 15 \%$ deficit on the red side below $\sim 2000 \ \kms$ (Fig. \ref{fig_ha_symm}). This only represents weak evidence for dust in the ejecta. The 2D spectra do, however, show a clear asymmetry with most emission on the blue side. While the observed asymmetrical emission in $\Ha$ is a  possible explanation the prevalence of blue emission is indicative of dust obscuration.  Because most of $\Ha$ arises outside the metal core, where the dust presumably is located, it is, however, less affected by dust in the metal core than e.g, the [O I] emission. 
There are also some indications of asymmetries from the [Ca II] line profile (Fig. \ref{fig6b}), consistent with that in $\Ha$. As already mentioned, this line, however, originates in the same region as $\Ha$. 

The most direct evidence from the line profiles at early epochs came from the evolution of the [O I] \wll 6300, 6364 lines \citep{Lucy1989}. Lucy et al. found a blueshift of the peak by $\sim 600$ \kms \ at the time of the peak velocity shift, $\sim 530$ days. Figure \ref{fig_stis_profiles} represents the last observation of this line before the spectrum becomes affected by the X-rays. We here note that the peak of the [O I] \wl 6300 line also in this late observation is blueshifted by $\sim 1000$ \kms. The resolution of this spectrum is limited, but the blueshift is consistent with that determined by \cite{Lucy1989}.

L13  argue that the 'hole' seen in the optical HST images is the result of the external X-ray illumination rather than dust obscuration. We can, however, not exclude that this region contains some dust. After all, this metal rich region is the most likely dust forming region. 
The VLT/SINFONI IFU spectra from 2005 and 2010 also show a smooth decrease of the red wing of the [Si I]/[Fe II] \wl 1.644 $\mu$m line at velocities $\la 3000$ \kms  \ (L13). This is roughly what is expected from dust internal to the line emitting region.

Summarizing this discussion, we find indications of dust from the 1999 [O I] \wl 6300 line (and also the  [Si I]/[Fe II]) profile, but also caution that the asymmetries in the core may also give this effect. From the high resolution observation of the H$\alpha$ line there is, however, little indication of dust, although this line should be less affected by this.

\section{Summary and Conclusions}
\label{sec_summary}
We have here discussed the spectral evolution of the ejecta and reverse shock emission from 2000 to 2012 (days $\sim 4400 - 9100$). Together with the HST imaging this represents a unique data set. Our main conclusions are summarised below:
\begin{itemize}
\item Both the H$\alpha$, [Ca II] and Mg II lines from the inner ejecta have increased by a factor of 4-6 from 2000 to 2012. This confirms the broad band increase found from the HST observations by L11.  There is a flattening of the light curve later than $\sim 7000$ days, possibly correlated with a similar flattening of the X-ray flux and the flux from the narrow lines from the shocked ring. 
\item The reverse shock flux behaves in a similar way. The leveling off of the flux is probably not a result of an increasing ionization of the pre-shock hydrogen, but may be a result of an arrival of protrusions of the inner, dense core region at the reverse shock, as seen from the HST imaging. 
\item The reverse shock emission extends to velocities $\ga 11,000$ \kms. This is larger than can be contained within the equatorial ring and indicates an anisotropic expansion of the ejecta above and below the ring plane.
\item The Balmer and metal lines have an inner ejecta component with a width of $\sim 4500$ \kms,  
\item From the last radioactively dominated STIS spectrum in 1999 (day 4381 -- 4387) we find similar velocities for the $\Ha$, Na I and [Ca II] lines, consistent with an origin in the hydrogen rich gas. The [O I] has a lower velocity, mainly coming from the core. 
\item There is from $\Ha$ and [O I] \wl 6300 evidence of a blueshift similar to what was seen as a dust indicator at early epochs. 
\item We identify a line feature at  $\sim$ 9220 \AA \ with an Mg II line pumped by Ly$\alpha$ fluorescence. The Mg II emission is mainly from primordial magnesium in the H and He rich ejecta. This line has been increasing faster than the other lines from the inner ejecta.
\item The deposition of the X-rays from the shock region affects mainly the hydrogen envelope and outer core region of the ejecta. Only X-rays with energy $\ga 5$ keV reach the inner core region. 
\item The changes in the morphology of the ejecta during the last decade is probably mainly a result of the increasing ionization and heating of the ejecta by the X-rays. The steep density gradient in the hydrogen envelope may lead to a more concentrated ionization to the reverse shock with time. 
\item The excitation by the X-rays is similar to the gamma-ray and positron deposition as long as the ionization is low. Most of the heating is balanced by far-IR fine structure and molecular lines, while the optical and near-IR lines are powered by non-thermal excitation and ionization.
\end{itemize}

Future observations are needed to confirm the evolution and follow the increasing ionization of the ejecta by the external X-rays.

\begin{acknowledgements} 
This work is supported by the Swedish Research Council and the Swedish National Space Board. 
Based on ESO observational programs
66.D-0589(A), 70.D-0379(A), 074.D-0761(A), 078.D-0521(A,B), 080.D-0727(A,B), 082.D-0273(A,B), 086.D-0713(A), 088.D-0638(A,C). Support for GO program numbers 02563, 03853,04445, 05480, 06020, 07434,
08243, 08648, 09114, 09428,10263, 11181, 11973, 12241 was provided by
NASA through grants from the Space Telescope Science Institute,
which is operated by the Association of Universities for Research in
Astronomy, Inc., under NASA contract NAS5-26555.
\end{acknowledgements}

{\it Facilities:} \facility{HST (STIS)}, \facility{VLT (FORS)}, \facility{VLT (UVES)}

\bibliographystyle{apj}
\bibliography{sn1987a_broad_lines_v12}

\end{document}